\documentclass[12pt,preprint]{aastex}
\usepackage{natbib,makeidx}
\usepackage{lscape}

\received{2002 January 31}
\begin{document}

\title{Spectroscopy of Dwarf H{\sc ii} Galaxies in the 
Virgo Cluster. I. Data, Chemical Abundances and Ionization Structure}


\author{Jos\' {e} M. V\'{\i}lchez}
\author{jvm@iaa.es}
\affil{Instituto de Astrof\'{\i}sica de Andaluc\'{\i}a (CSIC) \\
Apdo. 3004, 18080 Granada, SPAIN}
\and
\author{Jorge Iglesias-P\'{a}ramo}
\author{jorge@astrsp-mrs.fr}
\affil{Laboratoire d'Astrophysique de Marseille\\
Traverse du Siphon, F-13376, Marseille Cedex 12, FRANCE}


\begin{abstract}
Long-slit spectroscopy has been obtained for a sample of 22 blue dwarf 
galaxies selected in the direction of the Virgo Cluster as part of 
a larger sample of Virgo blue dwarf galaxies for which deep H$\alpha$ 
imaging has been collected. Most of the galaxies in the present sample 
appear classified as BCDs or dwarf Irregulars in the Virgo Cluster Catalog.  
Line fluxes, H$\beta$ equivalent width and extinction coefficients,
spatial emission profiles, ionization structure and physical conditions
are presented for the galaxies of the sample. Chemical abundances 
have been derived either using a direct determination of the electron 
temperature or after detailed examination of the predictions of different 
empirical calibrations. The oxygen abundance derived for the sample of 
Virgo dwarf galaxies span in the range 7.6 $\le$ 12 + log O/H $\le$ 8.9 , 
and the corresponding nitrogen to oxygen abundance ratio cover from 
values typical of low metallicity field BCD galaxies up to near solar.

\end{abstract}


\section{Introduction}

Within the study of the evolution of galaxies it appears now well established
that the environment is playing a significant role. From theoretical grounds, it
could be expected that galaxies located in high density regions such as clusters
of galaxies may suffer the effects of interactions, giving rise to a certain
degree of mass losses or mass redistribution. 
Also, after the interaction with the (hot) intracluster medium (ICM), the
gaseous component of galaxies may be affected by ram pressure stripping or
evaporation (Haynes et al. 1984). The relatively high density of the ICM
together with the higher probability of encounters in (rich) clusters may
produce selective losses of gaseous galactic material. In this respect, the
existence of a morphology-density relation is in the line of the observed
deficit of gas-rich dwarf galaxies in dense environments ({\it e.g.} Binggeli et
al. 1987). 

A direct environmental impact is expected to be observed in the activity of star formation 
in galaxies ({\it e.g.} Hashimoto et al. 1998; Iglesias-P\' aramo \& V\' \i lchez 1999). This 
fact is of particular relevance for the issue of galaxy evolution, owing to the strong implications 
that gas flows (in/outflows), as well as gas stripping and/or pressure confinement, may have for the 
study of the chemical evolution of galaxies. The observational results available  show 
a significant difference in the observable H{\sc i } content between cluster and field spirals, 
the Virgo ones being H{\sc i } deficient ({\it e.g.} Cayatte et al. 1990; Solanes et al. 2001). 
Other results obtained for the molecular component however, show that the cold molecular 
clouds located near the center of these galaxies, which accumulate large amounts of H$_2$ 
and CO, do not appear to disintegrate into the ICM (Kenney \& Young, 1988; Boselli et al. 1997). 
Since gas-rich dwarf galaxies are fragile systems, because they show low
mass surface densities and rotation velocities (Gallagher \& Hunter 1989; 1984), 
it is expected that the impact of the environment should be significant for these objects. 
However, Hoffman et al. (1988) studied a large sample of dwarf galaxies in 
Virgo and found that, on average, they were not more stripped than spirals.
The sole existence of these H{\sc i }-rich dwarf galaxies in such a hostile environment 
pose serious questions to be investigated further. 

Analysis of the spectroscopic properties of local and distant samples of emission-line 
galaxies belonging to environments of different density, show that galaxies in higher 
density regions may appear statistically associated to somewhat lower star formation rates 
(Hashimoto et al. 1998; V\' \i lchez, 1995). 
Similar studies of emission line galaxies located in or near voids, specially at lowest 
density, point to levels of star formation activity comparable to, or slightly higher 
than field galaxies ({\it cf.} Popescu et al. 1999; Grogin and Geller 2000). 
At this point, the question has arisen as to whether other fundamental properties 
of dwarf galaxies, such as their metal content, may feel the impact of their 
environment. It has been claimed that this can be the case of the so called tidal dwarf galaxies
(Duc and Mirabel 1999; V\' \i lchez 1999); nonetheless, the study of the global spectroscopic 
properties of a sample of star-forming dwarfs in different environments (V\' \i lchez 1995) 
led to the conclusion that the majority of them follow rather well a metallicity-luminosity 
relation. Some of the Virgo dwarfs studied in the latter work presented distinct 
spectroscopic properties and it was suggested they may hold a larger abundance though no 
definitive conclusion could be reached until more and better data are available. In the
present work new and better spectroscopic data have been collected for a sample of Virgo dwarfs.
A subsequent investigation has been undertaken in order to study their chemical evolution, 
and the impact of environmental effects. Their relative contribution to the metal enrichment 
of the ICM and to the overall metallicity-(mass)-luminosity relation will be presented in 
a forthcoming paper (V\' \i lchez et al. 2002 in preparation; paper II).

Several Virgo dwarf galaxies have been already studied spectroscopically (Gallagher \& Hunter 
1989; Izotov \& Guseva, 1989; Lee et al. 2000) or multirange ({\it e.g.} Brosch et al. 1998). 
The present work is aimed to produce a systematic spectroscopic study of a large sample 
of star-forming dwarf galaxies in the Virgo Cluster. Previous results for  dIrr in Virgo    
(e.g. IC~3475, Vigroux et al. 1986; UGC~7636,  Lee et al. 2000) and, more recently in A~1367 
(CGCG~97-073 and CGCG~97-079, Gavazzi et al. 2001, Iglesias-P\' aramo et al. 2002) have shown 
direct evidences that ram pressure stripping is at work for cluster dwarf galaxies. 

Finally, the location of our sample dwarfs in the Virgo cluster would allow the star forming 
regions to be spatially resolved and, since all share approximately the same distance, 
minimizing all the distance-dependent uncertainties in the analysis.
In section 2 of this paper we report on the observations and data obtained for the sample galaxies.
In section 3, the results derived for
the ionization structure, physical properties and chemical abundances are presented. 
The discussion and conclusions from the analysis of the results are presented in section 4.  
Further work including the implications of these results for the spectrophotometric evolution 
of the Virgo dwarf galaxy sample and a global discussion of their properties will 
be presented in paper II.

\section{Observations and Data Reduction}

Within an ongoing project of deep imaging of the BCD and 
Irregular galaxies in the Virgo Cluster Catalog (VCC; Binggelli et al. 1985),
a total sample of 22 objects (21 Virgo plus 1 background) most of them 
classified as blue compact dwarf galaxies, were selected for our spectroscopic 
study. All the galaxies were selected in the direction of the Virgo central field 
(sampling within a radius $\sim$ 5 degrees from M~87) for which our H$\alpha$ 
survey is being carried out (Boselli et al. 2002). The majority of the objects in 
the sample belong to VCC morphological classes BCD, Im/BCD; some of them appear 
classified as Im, Sm, and few other are quoted peculiar Sp/BCD, dS?/BCD? or amorphous   
({\it cf.} Gallagher and Hunter 1989). 

The observations were carried out at the Observatorio del Roque de los Muchachos 
(ORM,  La Palma) during two three-nights runs, using the 4.2-m William Herschel 
Telescope (WHT) with the ISIS double-arm spectrograph at the Cassegrain focus.
For each slit position two spectra, one in the red and one in the blue, 
were taken simultaneously using the corresponding dichroic.
Typical seeing values were around 0.$\arcsec$8 to 1$\arcsec$ throughout the runs, 
oscillating around 2$\arcsec$ during the second part of the 1994 run. 
During the first part of the 1993 run light high cirrus were present.
All the exposures during both runs were made as closer to the zenith as 
possible (airmasses were always $\leq$1.3 except in one case, VCC~428, 
observed at 1.5) in order to minimize any differential atmospheric 
refraction effects.

The 1280$\times$1180 EEV3 detector with  22.5-$\mu$m pixels was used
for the red arm, and a 1124$\times$1124 TEK1 with  24-$\mu$ pixels
for the blue arm.
The effectively observed detector area was windowed to 1230 and 1020 pixels 
in the spectral direction for the EEV3 and TEK1 detectors, respectively, 
and to 600 pixels along the slit for both detectors. The spatial scale was of 
0.335 $\arcsec$ pixel$^{-1}$ and 0.357 $\arcsec$ pixel$^{-1}$ for the EEV3 
and TEK1 detectors, respectively, giving a total slit length of approximately 
3.4 $\arcmin$.

The dichroic was set at an effective wavelength of $\lambda$ 5400 \AA \ (1993) 
and $\lambda$ 5700 \AA \ (1994) in order to separate the spectral ranges of the blue 
and red arms.
The 316R grating used for the red arm had 316 line mm$^{-1}$ yielding a
reciprocal dispersion of 60.4 \AA\  mm$^{-1}$. The 300B grating used for
the blue arm had 300 line mm$^{-1}$ yielding a reciprocal dispersion
of 62.2 \AA\ mm$^{-1}$. A similar effective spectral resolution of 
4 \AA \ FWHM was reached for both arms. 

An additional spectral range in the near infrared was observed for two 
of the galaxies, VCC~1699 and VCC~144 , centered at $\lambda $ 9100 \AA .
This observational set up allowed us a total spectral coverage which includes  
lines from [O~{\sc ii}] $\lambda$ 3727 \AA \  up to 
[S~{\sc iii}] $\lambda$ 9520 \AA .

The journal of observations is shown in Table~\ref{tab1}. For each galaxy the 
VCC number (Column 1), other name (column 2), morphological type \footnote{For 
a definition of the morphological types see Binggeli, Sandage \& Tamman (1985); 
Sandage (1961), Sandage \& Tammann (1981), Sandage \& Binggeli (1984)} (column 3), 
1950 coordinates (columns 4, 5), slit position angle (column 6), central wavelength 
(column 7), total exposure (column 8) and date of the observation (column 9) are 
presented.  

The data reduction was performed at the IAC using the standard
software package IRAF\footnote{Image Reduction and Analysis Facility, written and supported at
the National Optical Astronomy Observatoires}, following the standard procedure of bias
correction, flatfielding, wavelength calibration, sky subtraction and
flux calibration. The correction for atmospheric extinction was performed
using an average curve for the continuous atmospheric extinction at
the ORM (King 1985); flux calibration was achieved using repeated observations 
of the standards sp1446+259 and sp0642+021 from the La Palma spectrophotometric 
set (Sinclair, 1996) which were observed throughout both runs. 
The sky subtraction process was efficient for most of the spectra, though the 
proximity of the Na~{\sc i}~D feature (sky and Galactic) 
to the emission of He~{\sc i}~$\lambda$~5876 \AA \ in some spectra 
might contaminate the flux measured for this line. 

Figure~\ref{fig1} show the spatial profiles of the galaxies (line plus continuum 
flux) along the slit position, extracted from the 2-D spectra at the wavelength 
corresponding to the H$\alpha$ line for each galaxy. Many of the spatial profiles 
are consistent with a single central source of emission, though some 
galaxies present a rich spatial structure. This is the case of VCC~848 for which 
up to four emission peaks can be disentangled in the spatial profile. Three 
spectra were extracted, labeled (a), (b) and (c) in Figure~\ref{fig1} --given 
the low signal to noise (S/N) of the emission lines in the fourth peak--, in addition 
to the integrated spectrum.

In this work an integrated spectrum has been analyzed for each galaxy, extracted 
adding the flux in the set of spatial increments under the corresponding 
H$\alpha$ and H$\beta$ spatial profiles, in order to maximize the final S/N ratio. 
Representative spectra of all the galaxies of the sample are shown in
Figure~\ref{fig2}.

\section{Results}

\subsection{Line Intensities}

Line intensities were measured using the task SPLOT running on the IRAF environment, by 
marking to continuum points around each line to be measured and adding the total excess 
flux over the continuum level within the two points.
Errors were determined for each line by taking into account the 
poissonian error associated with the total number of counts in the line plus
its continuum, the dispersion (rms) of the nearby continuum, the effect of
background subtraction, as well as the error associated with the exact
placement of the local continuum. Independent measurements of the spectra 
were carried out in order to produce final error estimates.  We believe 
that, though the relative sources of error are of varying importance for 
each line, absolute errors in the flux of faint lines should be dominated 
by the continuum subtraction. We have not given a formal error for the absolute
flux calibration of the spectra; we estimate an average error to be in
the range 15--20$\%$. 

Each arm of the spectrograph was independently flux calibrated; and the matching 
of the corresponding extremes of the blue and red spectra was fairly good, except 
in the case of VCC~848 (1993 run) and VCC~1437, for which the flux match between 
the spectra of both arms was rather poor and they were rescaled.
The spectra of the galaxy VCC~848, which was independently observed (at slightly 
different positions) during the two runs, give results consistent within the errors.   

In order to minimize the error associated with the relative (to H$\beta$) flux 
calibration over the full wavelength range (especially for faint lines), 
whenever possible line ratios were derived with respect to the flux of the nearest 
Balmer line --- H$\beta$ or H$\alpha$ ---.

In Figure~\ref{fig2} all the spectra of the sample galaxies are shown. As illustrated 
in the Figure, the spectra do not fit into an homogeneous class. They cover a wide 
range in spectral properties, such as the equivalent width of the emission lines, the 
strength and shape of the underlying stellar continua as well as the strength of the Balmer 
lines in absorption and/or emission. As two illustrative examples of spectra of our sample   
we underline VCC~428, which shows the typical spectrum of an H{\sc ii} region with a 
very faint continuum and strong emission lines, and VCC~1179, which shows 
strong apparent absorption in the Balmer series and very faint emission lines.

Underlying Balmer absorption is clearly present in the spectra of many 
galaxies of the sample. In order to correct line fluxes from this effect, 
a selfconsistent procedure has been applied which performed a fit to the Balmer 
decrement taking into account the contributions of the extinction and the 
underlying absorption.
In this procedure the equivalent width in absorption was assumed to be the same  
for all the Balmer lines used (from H$\delta$ to H$\alpha$), as expected 
for young ionizing clusters. 
The correction for underlying absorption applied ranges from equivalent widths equal to 
zero ({\it e.g.} for VCC~1313, VCC~802) up to values around 9 \AA \ in some objects showing a 
strong continuum contribution ({\it e.g.} VCC~1179, VCC~135, VCC~1437).
This range of values of the H$\beta$ equivalent width in absorption were found to be 
enough to bring the observed Balmer ratios into the theoretical values, within the errors. 
All the spectra were corrected for reddening using the value of the Balmer decrement. 
The reddening coefficient, $C$(H$\beta$), was derived using the ratios of the
optical Balmer recombination lines once corrected from underlying absorption,
compared to the theoretical values for case B recombination (H$\alpha$/H$\beta$ $= 2.85$, 
H$\gamma$/H$\beta$ $= 0.469$ and H$\delta$/H$\beta$ $=0.260$, Hummer \& Storey 1987), 
and taking into account their S/N ratio and appropriate baseline corrections.
The values derived for $C$(H$\beta$) vary from  0 to 0.5; and a typical error for the 
reddening coefficient is estimated to be around $\pm$0.1 .

Measurements of the flux in the [O{\sc iii}]$\lambda$4363 \AA \ line were 
obtained from 9 spectra of the sample, for which a direct determination of the 
electron temperature was performed. Whenever possible, upper limits to the flux in 
the [O{\sc iii}]$\lambda$ 4363 \AA \ line were computed for each spectrum as 3 times 
the $\sigma$ of the flux observed at this wavelength. The electron temperatures 
derived from the two independent observations of VCC~848 are found to be fairly 
consistent within the errors.

Reddening-corrected (Whitford 1958) line intensities normalized to H$\beta$, 
are presented in Table~\ref{tab2} together with the derived flux of H$\beta$ 
(not extinction corrected), and the values of $C$(H$\beta$) and EW(H$\beta$) in emission   
for the sample galaxies. The errors quoted in Table~\ref{tab2}  
do not include the contribution from the uncertainty in the reddening coefficient. 

For the galaxies in common, we have compared our data with previous spectroscopic results 
in the literature (Gallagher and Hunter 1989, Izotov and Guseva 1989). We find general 
agreement for the values of R$_{23}$ derived, especially for the brighter objects: VCC~144, 
VCC~1725, VCC~1374, VCC~334. For VCC~2033 the match is somewhat poorer, as in the case of 
VCC~848; this seems to be largely a consequence of the different aperture
sizes and position angles of the slit used by the different authors.
It is important to point out here that, as shown in Table~\ref{tab3}, for the
objects in common, the three independent studies give values of R$_{23}$ which are 
consistent within the errors, though they were performed in different epochs
and with different telescope-spectrograph combinations.
In the case of VCC~72 (15$\deg$ 9) in the Virgo cluster catalog, its large radial 
velocity of 6351 kms$^{-1}$ (RC3 catalog; confirmed in our spectra) prevent us from 
including this galaxy in the abundance study of Virgo dwarfs.   

\subsection{Physical Conditions and Abundance Analysis}

Electron densities, $N_{\rm e}$, have been derived from the 
[S~{\sc ii}]$\lambda$$\lambda$ 6717/6731 \AA \ ratio using 
standard algorithms ({\it e.g.} McCall 1984; Aller 1984). All 
the regions studied give values typical of low-density 
H~{\sc ii} regions. For the objects in common this result is 
consistent with previous work (Gallagher and Hunter 1989; 
Izotov and Guseva 1989).

For those galaxies with [O{\sc iii}]$\lambda$4363 \AA \ flux measured, the electron 
temperature of the high ionization zone, $t_{[{\rm O III}]}$ was derived 
using FIVEL (Shaw \& Dufour 1995) assuming the low density limit. 
The temperature of the low ionization zone, $t_{[{\rm O II}]}$, 
has been determined assuming its relation with $t_{[{\rm O III}]}$ from 
photoionization models (Pagel et al. 1992). 
For the galaxies VCC~1699 and VCC~144 the [S{\sc iii}]$\lambda$6312 \AA \ line 
was measured; from the ratio of its intensity to the flux of the nebular [S{\sc iii}] 
lines a value of the  $t_{[{\rm S III}]}$ temperature has been obtained, which appears 
consistent with $t_{[{\rm O III}]}$ to within the errors.
Although the appropriate temperature for the [S{\sc iii}] and  [Ar{\sc iii}] 
lines is better represented by intermediate values between $t_{[{\rm O II}]}$ and 
$t_{[{\rm O III}]}$ (Garnett 1992), for the purpose of this paper we have 
used $t_{[{\rm O III}]}$ to derive their corresponding abundances.

For the galaxies with a direct determination of the electron temperature, 
the following ionic and total abundances and abundance ratios: O$^{++}$/H$^{+}$, 
O$^{+}$/H$^{+}$, O/H, N$^+$/O$^{+}$, S$^{++}$/H$^{+}$, S$^{+}$/H$^{+}$, S/H, 
Ne$^{++}$/O$^{++}$, Ar$^{++}$/O$^{++}$ 
have been computed using FIVEL (Shaw \& Dufour 1995) assuming the low density 
limit, and they are presented in Table~\ref{tab4}.
  
For those galaxies without a direct measurement of a temperature sensitive line, 
the total abundance of oxygen was derived using a sample of different existing 
abundance calibrations.
The behavior of optical oxygen line intensities in H{\sc ii} regions as a function of 
oxygen abundance has been studied via semiempirical methods (e.g. Pagel et al. 1979; 
Alloin et al. 1979; Edmunds \& Pagel 1984; Skillman 1989; Zaritsky et al. 1994; 
Pilyugin 2000) and also using theoretical photoionization models (e.g. 
Dopita \& Evans 1986; McGaugh 1991; Olofsson 1997; Dopita et al. 2000; 
Charlot \& Longhetti 2001). 

Several abundance calibrations can be found in the literature that make use of different 
ratios of relatively bright nebular lines. Among these empirical abundance parameters 
are R$_{23}$ (Pagel et al. 1979), P (Pilyugin 2000) and  $S_{23}$ (V\' \i lchez 
\& Esteban 1996). Additional empirical calibrations make use of the line ratios 
[O{\sc iii}]$\lambda\lambda$4959,5007/[N{\sc ii}]$\lambda\lambda$6548,84 (Alloin et al. 
1979; Dutil \& Roy 1999), [N{\sc ii}]$\lambda\lambda$6548,84/$H\alpha$ and 
[N{\sc ii}]$\lambda\lambda$6548,84/[O{\sc ii}]$\lambda$3727 (van Zee et al. 1998).

The optical oxygen lines [O{\sc ii}] and [O{\sc iii}] are sensitive to both oxygen abundance 
and electron temperature. At low metallicities, H{\sc ii} region cooling appears dominated 
by collisional excitation of H Lyman$\alpha$. Within this regime the total oxygen line 
intensity increases with abundance. At high metallicities, the oxygen abundance becomes 
the main coolant, and the cooling is then transferred from the optical to the infrared 
fine-structure lines [O{\sc iii}]$\lambda\lambda$52, 88$\mu$. Correspondingly, in this 
regime the intensities of the optical oxygen lines decline as the oxygen abundance increases. 

On this basis, the abundance parameter R$_{23}$, defined by Pagel et al. (1979) as R$_{23}$ = 
([O{\sc ii}]$\lambda$3727 + [O{\sc iii}]$\lambda\lambda$4959,5007)/H$\beta$ has been widely 
used in the literature as an abundance index. Also from this behavior, the 12 + log(O/H) 
versus R$_{23}$ abundance calibration results double valued, presenting an ambiguity between 
the high and low abundance branches, together with a turnover region centered between 
12 + log(O/H) $\approx$ 8.2-8.3 (e.g. McGaugh 1991, 1994; Miller \& Hodge 1996; Oloffson 1997). 
R$_{23}$ has been calibrated by different authors (Edmunds \& Pagel 1984; McCall et al. 1985; 
Dopita \& Evans 1986; see a comparison in McGaugh 1991). 
These calibrations might differ systematically at high-metallicity (i.e. log$R_{23}$ $<$ 0) 
by up to 0.2-0.3 dex, and there are some indications that empirical abundances in this range 
might be overestimated (e.g. Castellanos et al. 2002; Pilyugin 2001; Kinkel \& Rosa 1994; see 
also Bresolin \& Kennicutt 2002). The typical uncertainties quoted for R$_{23}$ empirical 
abundances are between 0.1 dex at low metallicity, up to 0.2 dex in the turnover region and 
the high abundance end. 

The parameter S$_{23}$, defined as S$_{23}$ = ([S{\sc ii}]$\lambda\lambda$6716,31 
+ [S{\sc iii}]$\lambda\lambda$9069, 9532)/H$\beta$ (V\' \i lchez \& Esteban 1996) has been 
calibrated by D\' \i az \& P\' erez-Montero (2000) as an empirical abundance index, in 
analogy to R$_{23}$. Spectroscopically, the sulphur lines defining S$_{23}$ are analogous 
to the optical oxygen lines in R$_{23}$ though, given their longer wavelengths, their 
contribution to the cooling become important at lower temperatures (D\' \i az \& 
P\' erez-Montero 2000). Thus, S$_{23}$ appear to show a monotonic behavior over a larger 
range of oxygen abundance. 
Kennicutt et al. (2000) have found that S$_{23}$ can show considerable scatter 
within an HII region and suggested it should be applied to integrated spectra; nonetheless 
the dispersion they derived in the abundances when using S$_{23}$ was only slightly larger 
than in the case of R$_{23}$. This calibration can provide more accurate abundances 
for objects with  oxygen abundances between 12 + log(O/H) = 7.20 (0.02 Z$\odot$) and 
12 + log(O/H) = 8.80 (0.75 Z$\odot$) (cf. D\' \i az \& P\' erez-Montero 2000).

Pilyugin (2000; 2001) has proposed a new empirical method for the determination of 
the oxygen abundance in H{\sc ii} regions based on the definition of the parameter 
P = ([O{\sc iii}]$\lambda\lambda$4959,5007/H$\beta$)/R$_{23}$.
Following earlier suggestions by McGaugh (1991) that the strong oxygen lines of [O{\sc ii}] 
and [O{\sc iii}] have enough information for the determination of abundances in H II regions, 
he compared oxygen abundances in bright H{\sc ii} regions derived from direct determination 
of the electron temperature, with those derived through the proposed P-method. He has found 
that the precision of the oxygen abundances derived with the P-method is comparable to the 
one obtained using a direct determination of the electron temperature. 
Two abundance relations were obtained: one for low metallicity HII regions, lower than 
12 + log(O/H)$\approx$ 8.1, and another one for high metallicity HII regions with 
12 + log(O/H) $\ge$ 8.2 (i.e. similar to the upper and lower abundance branches of the 
O/H vs. R$_{23}$ relationship).

In his theoretical work, McGaugh (1991) produced an extensive grid of H{\sc ii} region models
parameterized by the shape of the ionizing spectrum, the geometry of the nebula, and the 
abundance of the gas. He found that the behavior of the strong oxygen lines can be modeled 
by taking into account the softening of the ionizing spectra produced by stars of increasing 
metallicity. 
In these models, the ratio [O{\sc iii}]$\lambda\lambda$4959,5007/[O{\sc ii}]$\lambda$ 3727 
and R$_{23}$ are used to constrain both, the metallicity and the ionization parameter.  
The [O{\sc iii}]/[O{\sc ii}] vs. R$_{23}$ surface predicted by the models is double valued 
due to the two cooling regimes described above operating at high and low abundance. 
This theoretical calibration appears consistent with recent observations and 
photoionization models of high abundance objects (e.g. Castellanos et al. 2002 and 
references therein). It has an estimated mean uncertainty of 0.1 dex. Overall, we should 
bear in mind that the geometry and aging of the H{\sc ii} regions could introduce some 
scatter in the calibration (e.g. Stasinska 1999). 

Dopita et al. (2000) theoretically have recalibrated the extragalactic H II region 
sequence and defined the theoretical boundary between H{\sc ii} regions and active galactic 
nuclei. They used the PEGASE (Fioc \& Roca-Volmerange 1997) and STARBURST99 (Leitherer et al. 
1999) codes to generate the spectral energy distribution of the ionizing star clusters, and 
MAPPINGS3 (Sutherland \& Dopita 1993) was used to compute photoionization models, including 
a self-consistent treatment of dust physics and chemical depletion. The extragalactic H{\sc ii}
region sequence of observations is well reproduced by these models, assuming that the ionizing 
clusters are all young ($\le$ 2Myr) (these authors point out this is likely a selection effect).
They proposed that the ratio [N{\sc ii}]$\lambda\lambda$6548,6584/[O{\sc ii}]$\lambda$3727 gives 
the best diagnostic of abundance, being monotonic between 0.1 Z$\odot$ and over 3.0 Z$\odot$.

More recently, Charlot and Longhetti (2001) combined recent population synthesis and photoionization 
codes to compute the line and continuum emission from star-forming galaxies. They calibrate the 
nebular properties of the models using observations of line ratios for a sample of star-forming 
galaxies. From optical spectral fits they are able to constrain the star formation history, the 
gas abundance, and the absorption by dust. 

For the determination of abundances in this paper we have used: 
i) the analytical expression of the McGaugh (1991) theoretical models reported in Kobulnicky et al. 
(1999); ii) the theoretical calibration of [N{\sc ii}]/[O{\sc ii}] after Dopita et al. (2000); 
together with iii) the empirical calibration of S$_{23}$ (D\' \i az \& P\' erez-Montero 2000);  and 
iv) the empirical calibration of the parameter P (Pilyugin 2000; 2001). For each galaxy, an 
average oxygen abundance $<$12+log(O/H)$>$, and its N/O  abundance ratio, $<$log(N/O)$>$, have 
been finally adopted. In the next section we discuss the abundance analysis and present the 
final adopted abundances for the sample galaxies. 
 
\section{Discussion and Conclusions}

In this work oxygen abundances and N/O abundance ratios have been derived for 21 dwarf 
galaxies in the Virgo Cluster. These results show that the oxygen abundances of the galaxies 
span in the metallicity range 0.04 $Z\odot$ $\le$ Z $\le$ 1 Z$\odot$. The nitrogen to oxygen 
ratios derived ranging from values typical of low metallicity, field BCD galaxies up to 
near solar ratios. The oxygen abundance and the nitrogen to oxygen abundance ratios for our 
Virgo sample are presented in Table~\ref{tab5}. Comparison with previous work on abundances 
of Virgo dwarfs is a difficult task due to the shortage of papers devoted to this subject.   
Previous spectroscopy of several star-forming Virgo dwarfs has been done by Gallagher \&  
Hunter (1989) and Izotov \& Guseva (1989). Bright line ratios, such as R$_{23}$, can be 
derived for some of their objects allowing, in principle, a rough estimation of the abundance 
to be done. Unfortunately, given the rather poor S/N of the spectra, the faint [N{\sc ii}] 
lines were not available in previous works, implying a fairly large uncertainty. 
Only an average value of the oxygen abundance of 12+log(O/H)$\approx$ 8.2 $\pm$ 0.3, 
is available in Gallagher \& Hunter (1989) for their sample of Virgo dwarfs. This 
abundance was derived using the R$_{23}$ calibration by Edmunds \& Pagel (1984) and
is consistent with the range of oxygen abundances derived in this work.
For the four galaxies in common with the sample of Izotov \& Guseva (1989):
VCC~144, VCC~324, VCC~1437 and VCC~2033, the abundances they derived are between 
8.3 $\le$ 12+log(O/H) $\le$ 8.5 with a typical 1$\sigma$ error of 0.26 dex. They have 
derived abundances using an estimation of the electron temperature for each object 
using the empirical relation presented in Pagel et al. (1979). These abundance values 
appear consistent, within the errors, with the values derived in this work for the  
same objects. 

In order to adopt the oxygen abundance, $<$12+log(O/H)$>$, and nitrogen to oxygen 
abundance ratio, $<$log(N/O)$>$, for each galaxy of the sample, we have proceed 
as follows:
i) For the (9) objects with an electron temperature measurement, the abundances derived 
using the electron temperature, quoted in Table~\ref{tab4}, have been adopted and they 
are presented in Table~\ref{tab5} (In the following, we will refer to these abundances 
as ``direct'' abundances).  

ii) For each of the (13) remaining objects, an average abundance value, or the corresponding 
abundance interval, has been finally adopted and is presented in Table~\ref{tab5}. 
This average value for each galaxy was derived from the predictions of the corresponding 
abundance calibrations used in this work (see Section 3) as it is described below. In this respect, 
the dispersion among the abundance predictions for each object can be seen as an indicator 
of the uncertainty in the derivation process. We believe that the advantage of using several  
calibrations, instead of choosing a single one, is that we can obtain a more realistic 
estimate of the uncertainty (e.g. Zaritsky et al. 1994).      

At this point it is important to emphasize the role that the [N{\sc ii}]/[O{\sc ii}] ratio 
plays in the abundance calibrations. This line ratio has been shown to be monotonic with 
oxygen abundance between 0.2 Z$\odot$ up to 3.0 Z$\odot$ (Dopita et al. 2000), a metallicity 
range  where nitrogen behaves as a secondary element. 
It is also important to bear in mind that the [N{\sc ii}]/[O{\sc ii}] ratio depends on both, 
electron temperature and N/O abundance ratio. Thus, choosing a different formula accounting for 
the dependence of N/O on O/H in principle would lead to different values of the line ratio 
(e.g. Bresolin et al. 2002 and references therein). For example, Stasinska et al. (2001) models 
have adopted a mild dependence of N/O on O/H abundance (log(N/O) = 0.5log(O/H)+0.4) leading to 
smaller [N{\sc ii}]/[O{\sc ii}] ratios. 

Nevertheless, as originally proposed by Edmunds \& Pagel (1978), N/O may also change 
with time as a consequence of delayed production of nitrogen. This fact may have some 
relevance in the process of abundance determination, when comparing with the predictions 
for the zero-age case. Indications of such behavior have been pointed out on the plot of N/O 
versus O/H for H{\sc ii} regions and star-forming galaxies (e.g. van Zee et al. 1998; 
Henry et al. 2000) notably for abundances between 0.1Z$\odot$ to 0.5Z$\odot$. 
On the other hand, variations in the N/O ratio can also be produced due to the effects 
of gas inflow or outflow in the galaxy and/or self-enrichment (see Henry et al. 2000).  
For primary nitrogen, N/O can be affected only in the case of a differential outflow 
(i.e. different outflow for oxygen than for nitrogen); when nitrogen is secondary, 
N/O can be affected by an unenriched inflow in addition to the differential outflow
(Henry et al. 2000; Koppen and Edmunds 1999, and references therein).

The [N{\sc ii}]/[O{\sc ii}] ratio presents a single parameter sequence from high 
to low abundances (e.g. van Zee et al. 1998; Dopita et  al. 2000). As commonly 
accepted, in this work we have assumed that those H{\sc ii} regions showing 
log([N{\sc ii}]/[O{\sc ii}])$<$ -1.0 present lower oxygen abundances, with respect 
to those regions with log([N{\sc ii}]/[O{\sc ii}])$\ge$-1.0 which correspond to the high 
abundance regime (e.g. McGaugh 1994; Miller \& Hodge 1996; van Zee et al. 1998; Contini 
et al. 2002). In Table~\ref{tab5} the value of log([N{\sc ii}]/[O{\sc ii}]) has been
quoted for each galaxy, together with the parameters log([O{\sc iii}]/[O{\sc ii}]) and
log R$_{23}$. These line ratios are necessary to compute the abundance predictions from 
the different calibrations used in this work, as described in Section 3.

For each galaxy, in Table~\ref{tab5} we show the abundance predictions for 12+log(O/H) 
that have been obtained using the different calibrations. In the Table P$_{upper}$ and  
P$_{lower}$ refer to the lower and upper oxygen abundance, respectively, predicted from 
the empirical calibration of Pilyugin (2000; 2001); R$_{23}{_lower}$ and R$_{23}{_upper}$ 
are respectively the lower and upper oxygen abundance predicted by McGaugh (1991) theoretical 
models, computed using the algorithms reported in Kobulnicky et al. (1999); the abundances 
obtained from the theoretical calibration of the [N{\sc ii}]/[O{\sc ii}] diagnostic in 
the models of Dopita et al. (2000) are presented; and finally, the abundances predicted by 
the empirical calibration of S$_{23}$ after D\' \i az \& P\' erez-Montero (2000) are also quoted. 
We have quoted in the Table all the abundance predictions of the different calibrations for all 
the objects. However, in the computation of the average abundance only those predictions 
within the range of application of each calibration (sect. 3) were considered.

As an additional constraint, for the spectra with appropriate S/N, lower limits to O/H 
were computed using the measured upper limits to the [O{\sc iii}]$\lambda$4363 \AA \ flux. 
The lower limits to O/H are quoted also in Table~\ref{tab5}. This constraint was fulfilled, 
whenever meaningful, providing information on the abundance regime.

In Figure~3 we present the diagnostic diagram of log([O{\sc iii}]/[O{\sc ii}]) versus 
log([N{\sc ii}]/[O{\sc ii}]) with the points of our sample of Virgo galaxies; superposed  
is the models grid of Dopita et al. (2000) for abundances 0.2Z$\odot$ up to 2Z$\odot$.
Triangles and circles correspond to objects with log([N{\sc ii}]/[O{\sc ii}])$<$-1.0 and 
$\ge$-1.0 respectively. Filled points represent those objects for which a ``direct'' 
determination of the abundance has been obtained. A sequence of points can be seen in the
Figure, where the highest abundances, between 1 to 1.5Z$\odot$, are predicted for four 
objects in the sample.
Many points cluster around the 0.5Z$\odot$ line, while only three points --spanning the 
whole range in ionization parameter, log u, in the plot-- are consistent with the lowest 
abundance predictions shown for 0.2Z$\odot$ (according to Figure~6 in Dopita et al. (2000)
these points can be consistent also with the locus of the models for 0.1Z$\odot$). 
A comparison between the abundances predicted from these models and the ``direct'' abundances 
for the objects in Table~\ref{tab4} tell us that models give abundances which are consistent, 
within the errors, for the three objects with 12 + log(O/H)$\ge$8.2; the galaxies with ``direct'' 
abundances lower than 0.2Z$\odot$ are out of the range allowed for these models. 
 
The log([O{\sc iii}]/[O{\sc ii}]) versus log R$_{23}$ diagnostic diagram is presented in 
Figures~4 and ~5. McGaugh (1991) models, as reported in Kobulnicky et al. (1999), are 
shown for the lower, Figure~4, and higher, Figure~5, abundance regimes. The points are 
coded as in Figure~3. Model predictions for the oxygen abundance, increasing by 0.1 dex from 
12 + log(O/H) = 7.4, are shown in Figure~4 as continuous lines. In Figure~5, corresponding 
models are shown from 12 + log(O/H) = 8.2 by dotted lines, each line increasing by 0.1 dex.
As stated before, triangles represent objects which are to be analyzed using the low 
abundance models, whereas circles represent objects corresponding to high abundance models. 
The somewhat larger uncertainty expected in the turnover region, pointed out in Section 3,  
is apparent when extrapolating model predictions around 12 + log(O/H)= 8.2 in both figures. 
Looking at Figures~4 and ~5 and Table~\ref{tab5} we can see that, for most of the objects with 
a  ``direct'' oxygen abundance, these abundances are consistent with model predictions, within 
the nominal error of 0.2 dex. It is striking the large discrepancy between 
model's and ``direct'' abundances found for VCC~802, VCC~848 and VCC~1725. For these three 
galaxies R$_{23}{_upper}$ has been selected since they show log([N{\sc ii}]/[O{\sc ii}])$\ge$-1.0 . 
However, it is interesting to note that these discrepancies would be minimized adopting 
R$_{23}$${_lower}$ instead of R$_{23}$${_upper}$, under the hypothesis that [N{\sc ii}]/[O{\sc ii}] 
for these galaxies is higher than the value expected for dwarf galaxies with similar oxygen abundance. 

A comparison between ``direct'' and empirical abundances derived using Pilyugin (2000; 2001) 
for the galaxies in Table~\ref{tab4} give a smaller difference, typically comparable to he 
nominal error of the calibration. From Table~\ref{tab5} we can notice again 
the striking cases of VCC802, VCC~848, VCC~1725 for which we find that 
[12 + log(O/H) - P$_{upper}$] = -0.56, -0.35, -0.66 dex respectively. As in the case of the 
R$_{23}$ McGaugh's models noted above, if we would assume P$_{lower}$ instead of P$_{upper}$ 
for the three galaxies, these differences would decrease to [12 + log(O/H) - P$_{upper}$] = 
-0.06, -0.05, -0.26 dex, giving values closer to the nominal error of the calibration. 
In the case of VCC~1699 we find [12 + log(O/H) - P$_{lower}$] = 0.45 dex whereas 
[12 + log(O/H) - P$_{upper}$] = -0.06 dex. In this case, since VCC~1699 shows log R$_{23}$ $>$ 0.9, 
it is likely that this point belongs to the turnover region of the calibration and then, as 
described below, its expected empirical abundance would be consistent with its ``direct'' abundance.
 
Both, the R$_{23}$ and P abundance calibrations, when log R$_{23}$ is larger than 0.9 
enter a region around 12 + log(O/H)=8.2 to 8.3 where abundance predictions are 
more uncertain (see Sect. 3). This region corresponding to the turnover between low and high 
abundance branches of both calibrations. At this point, the [N{\sc ii}]/[O{\sc ii}] 
calibration from the theoretical models of Dopita et al. (2000) begins to work up to 
oversolar values. Nevertheless, these models for 0.1Z$\odot$ and for 0.2Z$\odot$ predict 
nearly the same locus in the log([N{\sc ii}]/[O{\sc ii}]) versus log([O{\sc iii}]/[O{\sc ii}]) 
diagram (Figure~6 in Dopita et al. 2000). This fact imply, in practice, that objects with 
abundances between 7.9 $\ge$ 12 + log(O/H) $\ge$ 8.2 would not be discriminated.
Thus, we have adopted a somewhat conservative approach assuming an average abundance 
of 12 + log(O/H)$\approx$8.2, with a 0.2 dex error bar, for those objects in the sample 
without a direct abundance determination and presenting logR$_{23}$ larger than 0.9. 

The S$_{23}$ parameter seems to be monotonic across this abundance region, but it was available 
only for two of the objects of the Virgo sample: VCC~144 and VCC~1699. For these two galaxies,
as can be seen in Table~\ref{tab5}, the oxygen abundances provided by the S$_{23}$ calibration 
(D\' \i az \& P\' erez-Montero 2000) are in good agreement with the ``direct'' abundances, 
within the nominal error of the calibration.  

For some of the objects without a ``direct'' abundance we have adopted an interval in 
12 + log(O/H) with the aim of encompassing the range in O/H predictions (typically between 
the McGaugh's and Pilyugin's calibrations). 

The average N/O abundance ratio, $<$log(N/O)$>$, has been derived for each galaxy of the 
sample. In doing that, an ``average'' electron temperature was assumed for each object, 
corresponding to the temperature for which the derived oxygen abundance can be reproduced 
from the observed [O~{\sc iii}] and [O~{\sc ii}] fluxes within the errors. 
From photoionization models, it is well known that for temperatures $t_{[{\rm O III}]}$ 
$\ge$ 1.2 K, $t_{[{\rm O III}]}$ is higher than $t_{[{\rm O II}]}$ and the opposite is 
found for $t_{[{\rm O III}]}$ below this value (e.g. Peimbert 2002 and references therein).
Using an average temperature would introduce an error when deriving the nitrogen to oxygen 
ratio (e.g. van Zee et al. 1998). Assuming a typical difference of 1.000 K between 
$t_{[{\rm O III}]}$ and $t_{[{\rm O II}]}$ within the average temperature range used, would 
imply a typical error of 0.06 dex in $<$log(N/O)$>$, which we believe is within the nominal 
error budget of the calibrations. 

The use of strong line diagrams, notably at high abundance, is subject to large uncertainties 
(see Stasinska 1999 for a review). McGaugh's (1991) models assume unevolving ionizing stellar 
clusters, whereas it is known that both, the ionization parameter as well as the shape of the 
ionizing spectrum may evolve (e.g. Stasinska \& Leitherer 1996; Olofsson 1997). The effect 
of cluster evolution on the oxygen line fluxes is expected to be important especially at high 
metallicity, where diagnostic diagrams could change as a function of time (e.g. Stasinska 1999). 
On the other hand, using the best empirical fit to the samples of abundances derived from 
temperature-sensitive lines, as in Pilyugin's P calibration, could lead to lower values 
of oxygen especially at high abundance (e.g. Stasinska 2002). 
    
There have been some suggestions that the R$_{23}$ calibration may provide abundances 
systematically higher by some 0.2-0.3 dex in the high abundance regime, probably depending on the
excitation conditions (e.g. Castellanos et al. 2002; Pilyugin 2001; Kinkel \& Rosa 1994; see 
also Bresolin \& Kennicutt 2002). The effect of excitation was been taken into account by 
McGaugh (1991), Pilyugin (2000) and Dopita et al. (2000). For two of the objects showing 
[N{\sc ii}]/[O{\sc ii}] $\ge$ 0.1, their errors could allow the lower branch to be chosen at 
the one sigma level, in these two cases, VCC~562 and VCC~2033, the lower branch of 12 +log(O/H) 
has been also indicated in Table~\ref{tab5}.

For all the objects with 12 + log(O/H) $\le$ 8.2 the nitrogen to oxygen ratios derived 
present values typical of low abundance dwarf galaxies. There are two notable exceptions to 
this behavior, which deserve independent confirmation and further study: VCC~1725 and VCC~802. 
Our data for these two objects indicate oxygen abundances below 0.1 solar but their N/O values 
appear somewhat higher, close to the Orion value of log(N/O) = -0.85 $\pm$ 0.10 (Esteban et al. 1998). 
Among the galaxies with 12 + log(O/H) $\ge$ 8.2 , a significant fraction present high abundances 
with O/H and N/O near to or larger than the solar values of 12 + log(O/H) = 8.71 $\pm$ 0.05 
and log(N/O) = -0.78 $\pm$ 0.12 (Holweger 2001; Allende Prieto et al. 2001). 

Finally, near 50$\% $ of the galaxies of the sample present oxygen abundances in between the 
abundance of the Large Magellanic Cloud, 12 + log(0/H) = 8.39 $\pm$ 0.12 (Pagel et al. 1978), 
and that of Orion, 12 + log(0/H) = 8.64 $\pm$ 0.06 (Esteban et al. 1998). Most of the galaxies 
presenting the larger O/H abundances or higher N/O ratios tend to be associated with spectra 
showing an important continuum contribution and conspicuous absorption lines. These outstanding 
objects deserve further study; in particular, following the suggestion that the N/O ratio could 
indicate the time since the bulk of star formation has occurred (Edmunds \& Pagel 1978; Matteucci 
\& Tosi 1985), those objects presenting the highest nitrogen to oxygen ratio we believe could be 
good candidates to host post-starburst galaxies. Therefore these objects may provide 
interesting hints for chemical evolution modeling of dwarf galaxies in dense environments. 

A detailed study of the results obtained in this work will be presented in paper~II. 
In the framework of the metallicity-(mass)-luminosity relation (Skillman et al. 1989; 
Richer et al. 1998; Hidalgo-G\'{a}mez \& Olofsson 1998; Pilyugin 2001) paper~II is aimed 
to analyze the relationship between metallicity and the fundamental properties and chemical 
evolution of Virgo dwarf galaxies.

\acknowledgments
The WHT is operated on the island of La Palma by the ING in the Spanish Observatorio del 
Roque de Los Muchachos of the Instituto de Astrof\'{\i}sica de Canarias. We acknowledge CAT 
for the allocation of telescope time to this project. We thank an anonymous referee 
for many interesting suggestions and comments which have helped to improve this paper.
JMV thanks the I.A.C. (Tenerife) and the L.A.S. (Marseille) for hospitality and partial funding 
during the realization of this work. JMV acknowledges financial support from a CSIC ``Marina Bueno'' 
grant. JIP acknowledges a European ``Marie Curie'' Fellowship. This research is supported by 
project AYA2001-3939-C03-01 of the Spanish Plan Nacional de Astrom\' \i a y Astrof\' \i sica.


\clearpage

\begin{figure}
\epsscale{0.90}
\plotone{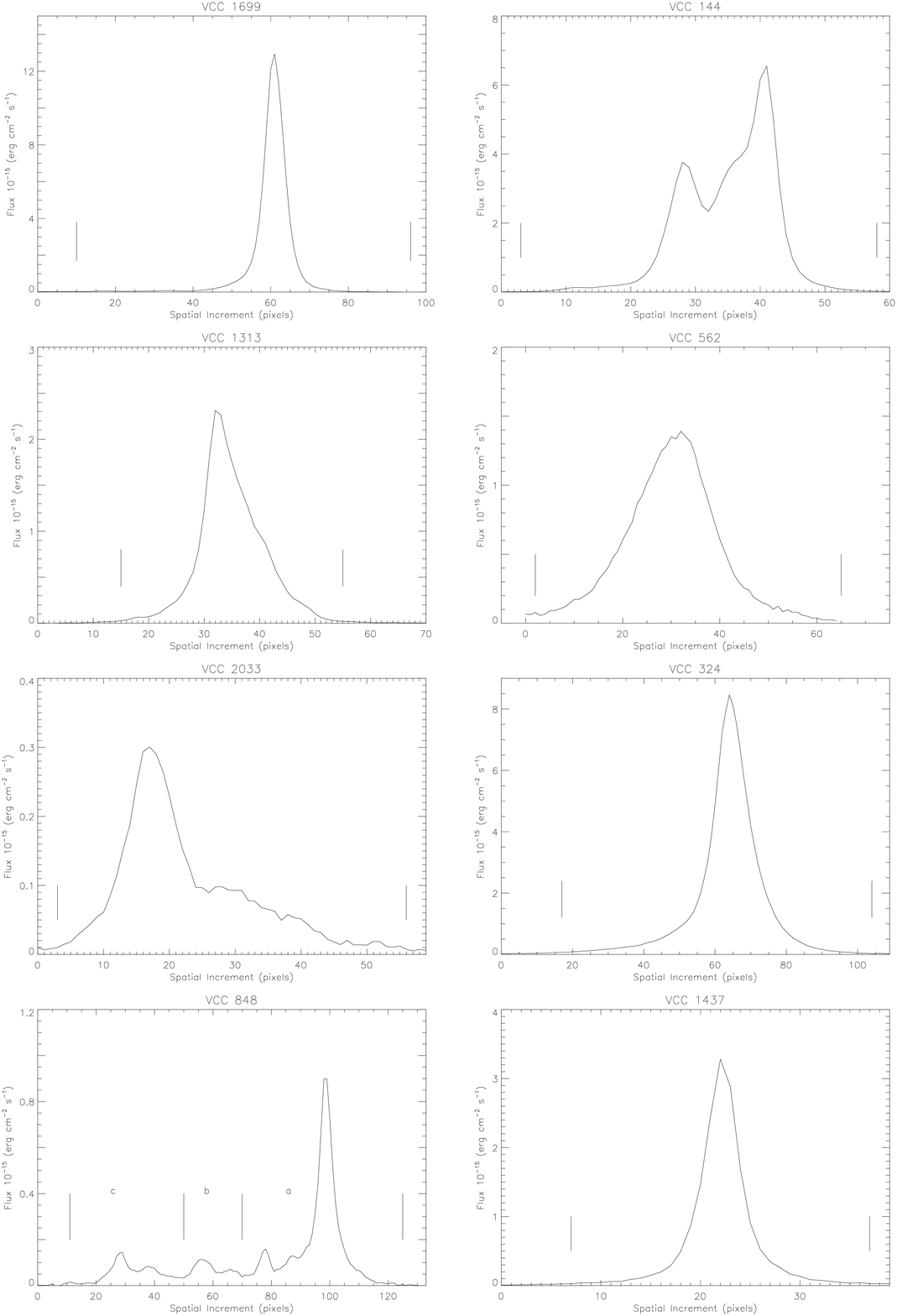}
\caption{Spatial emission profiles of the sample galaxies in a wavelength
    window centered in H$\alpha$. The spatial scale is 0.335 $\arcsec$ pixel$^{-1}$.}
\label{fig1}
\end{figure}

\addtocounter{figure}{-1}

\begin{figure}
\epsscale{0.90}
\plotone{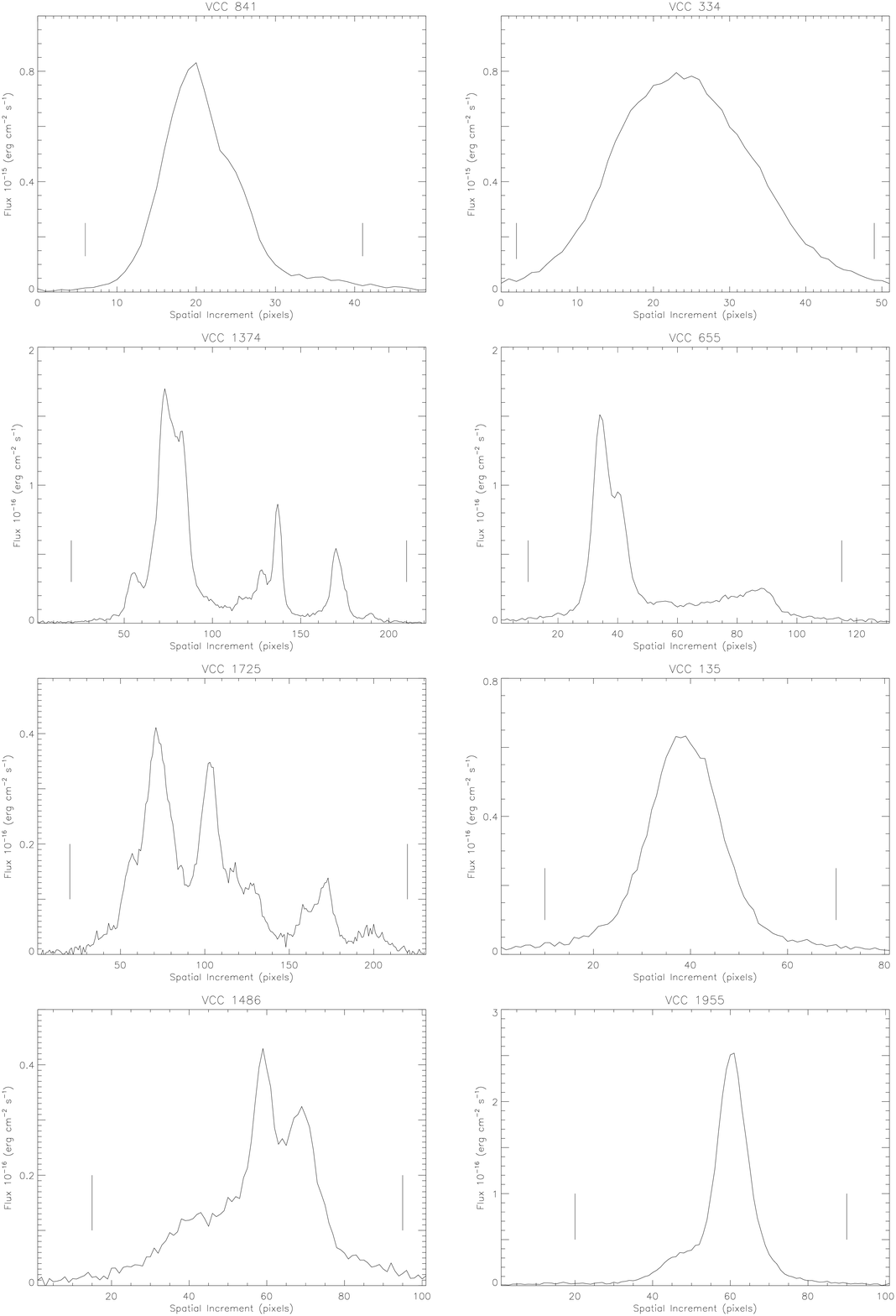}
\caption{Continued.}
\label{fig1}
\end{figure}

\addtocounter{figure}{-1}

\begin{figure}
\epsscale{0.90}
\plotone{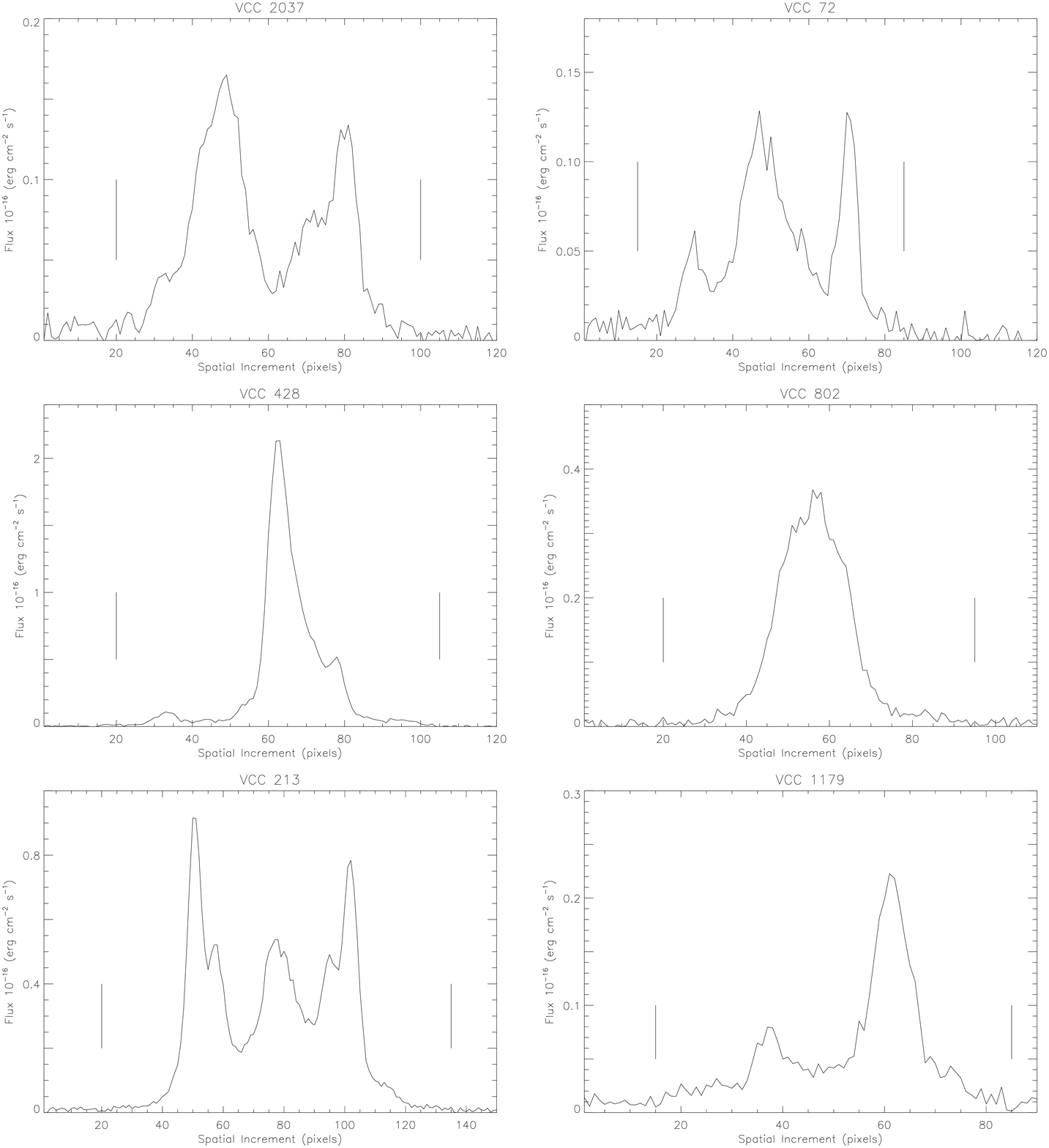}
\caption{Continued.}
\label{fig1}
\end{figure}

\clearpage
\begin{figure}   
\epsscale{0.90}
\plotone{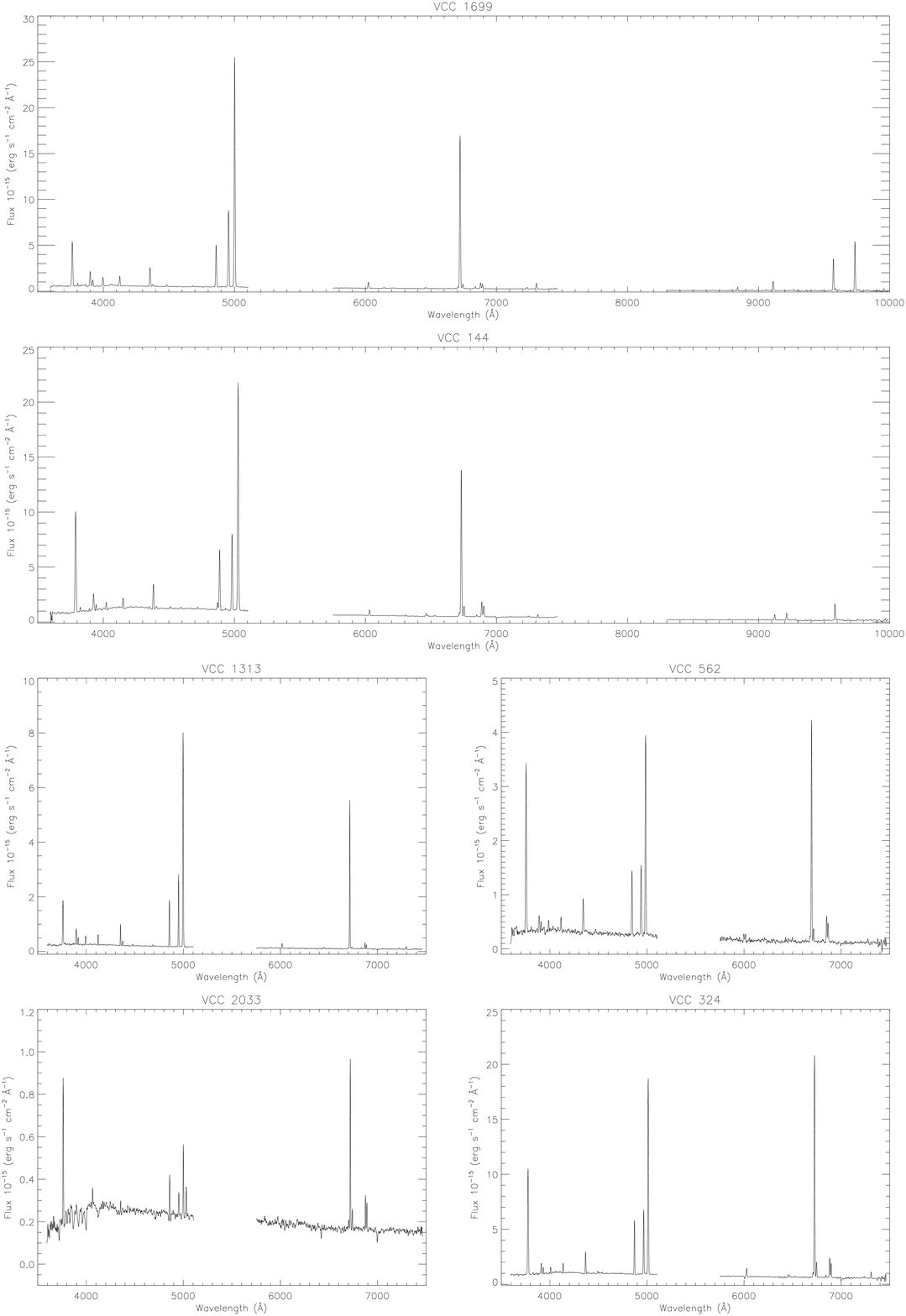}
\caption{Spectra of the galaxies of the Virgo sample.}
\label{fig2}
\end{figure}
\addtocounter{figure}{-1}
\begin{figure}   
\epsscale{0.90}
\plotone{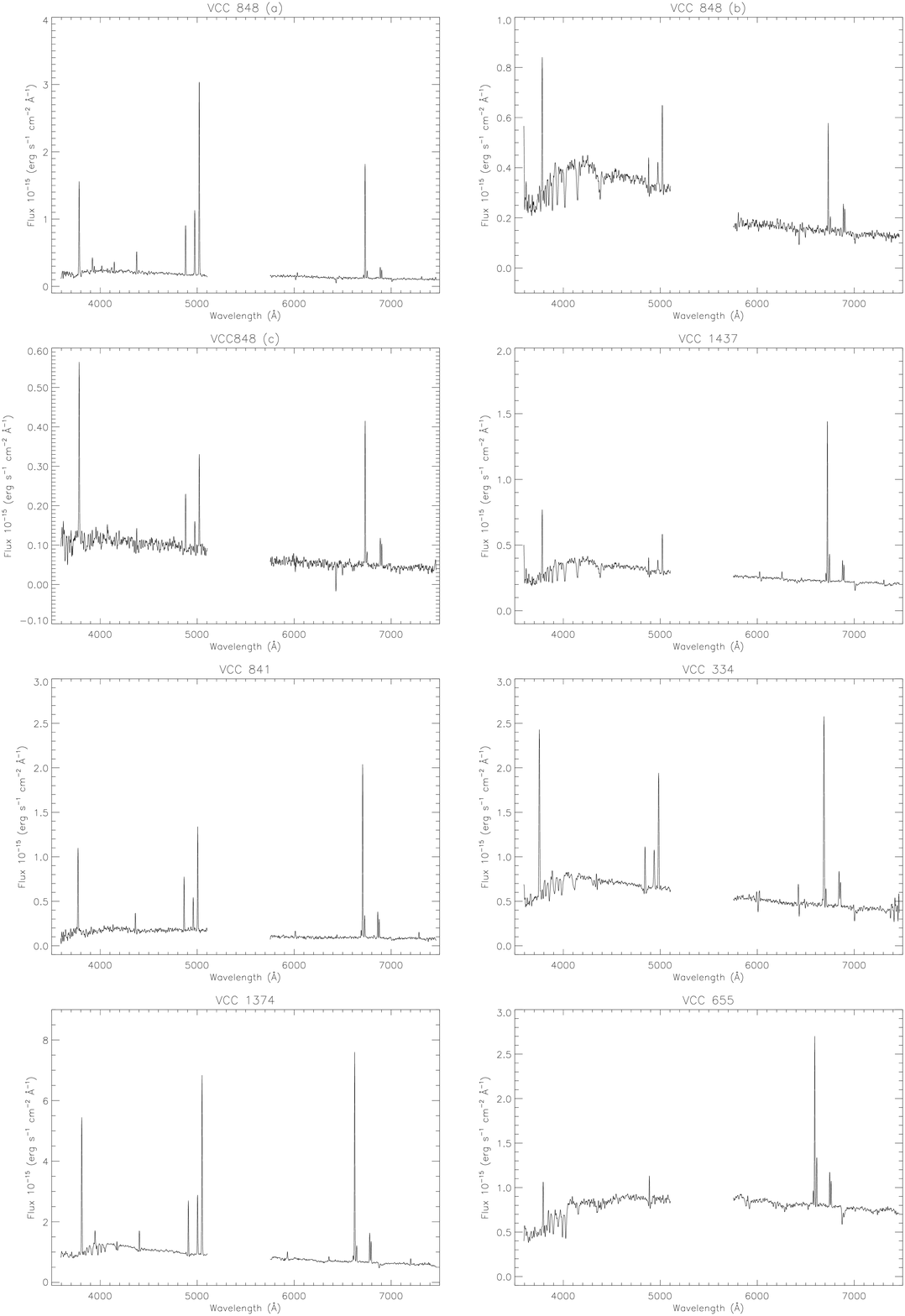}
\caption{Continued.}
\end{figure}
\addtocounter{figure}{-1}
\begin{figure}   
\epsscale{0.90}
\plotone{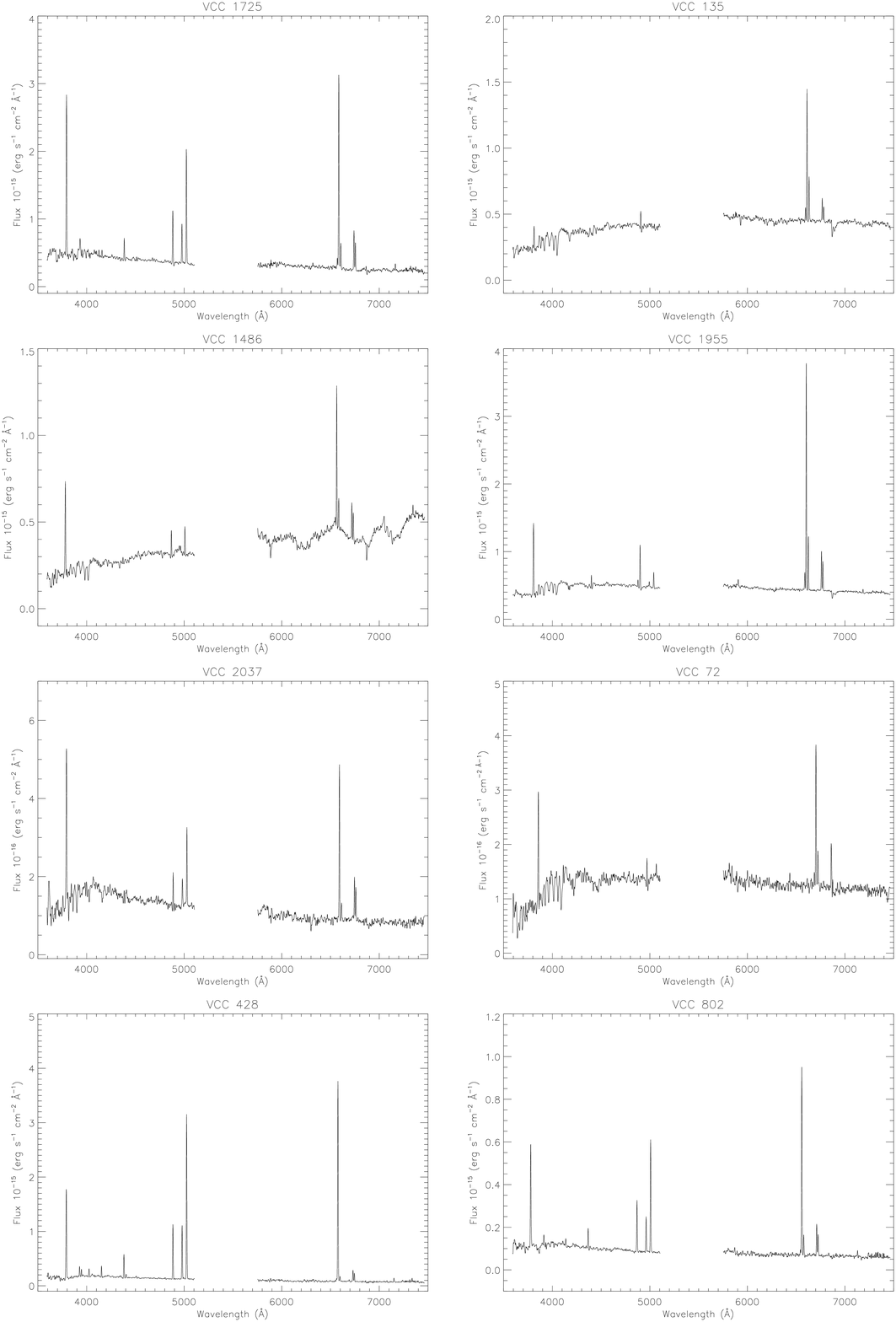}
\caption{Continued.}
\end{figure}
\addtocounter{figure}{-1}
\begin{figure}   
\epsscale{0.90}
\plotone{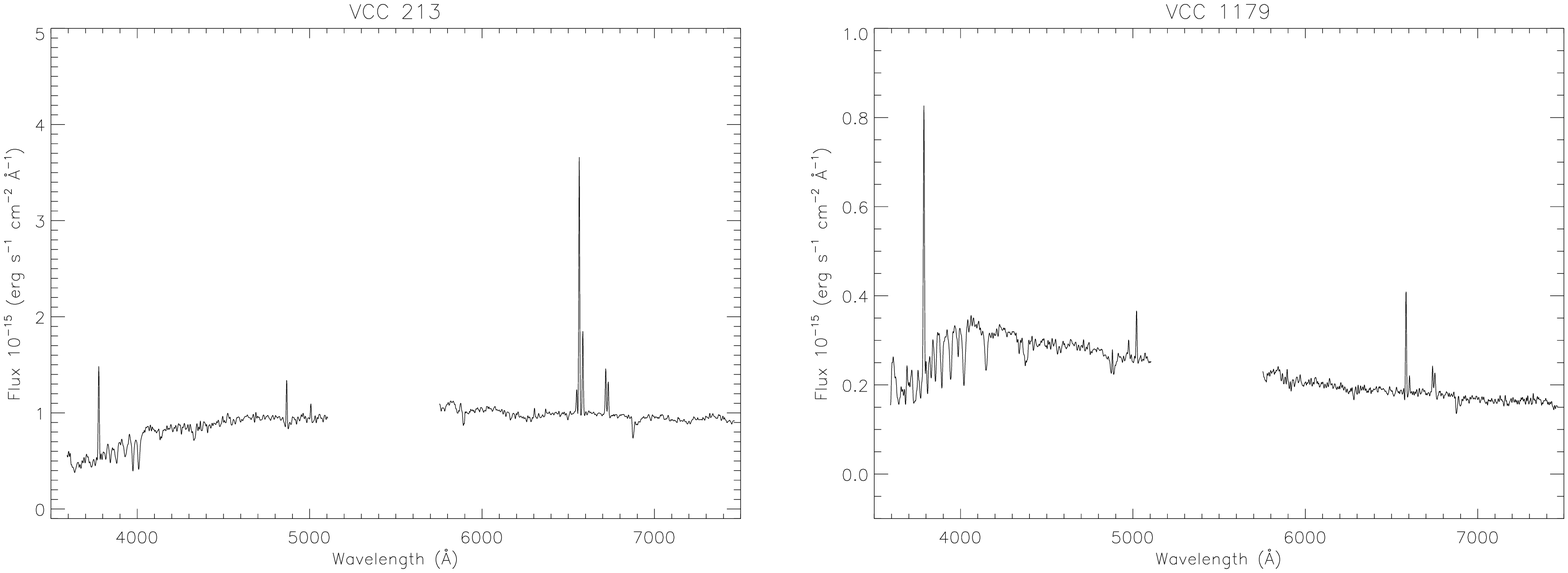}
\caption{Continued.}
\end{figure}

\clearpage
\begin{figure}   
\epsscale{1.0}
\plotone{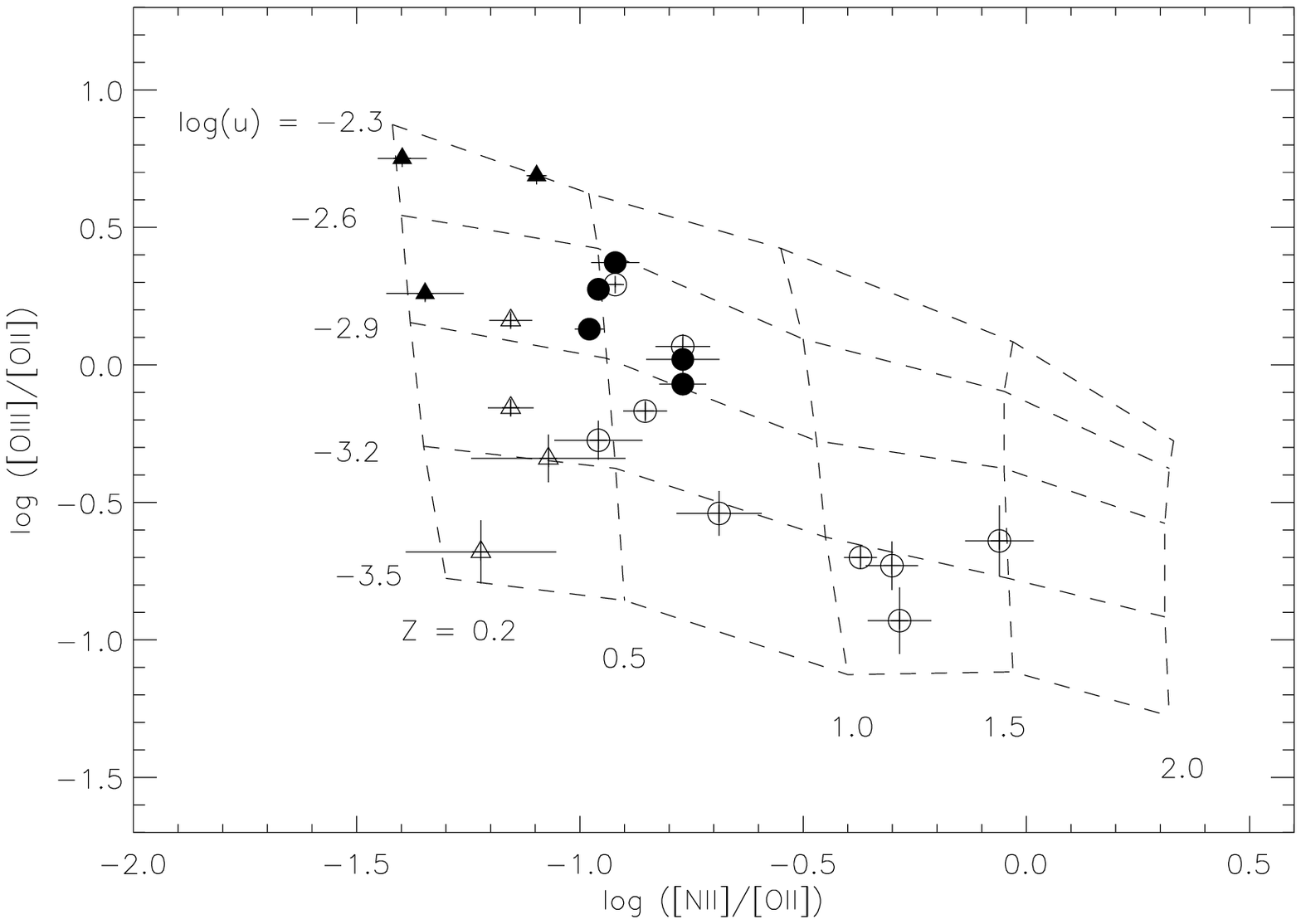}

\caption{Diagnostic diagram log([O{\sc iii}]/[O{\sc ii}]) vs. log([N{\sc ii}]/[O{\sc ii}]) with
the model grid of Dopita et al. (2000) superposed for abundances 0.2Z$\odot$ up to 2Z$\odot$.
Points represent the Virgo galaxies studied here coded as follows: triangles correspond to 
objects with log([N{\sc ii}]/[O{\sc ii}])$<$-1.0; circles correspond to objects with 
log([N{\sc ii}]/[O{\sc ii}]) $\ge$-1.0. Filled points represent objects with a ``direct'' 
determination of abundance. See the text for details.}
\label{fig3} 
\end{figure}

\clearpage
\begin{figure}   
\epsscale{1.0}
\plotone{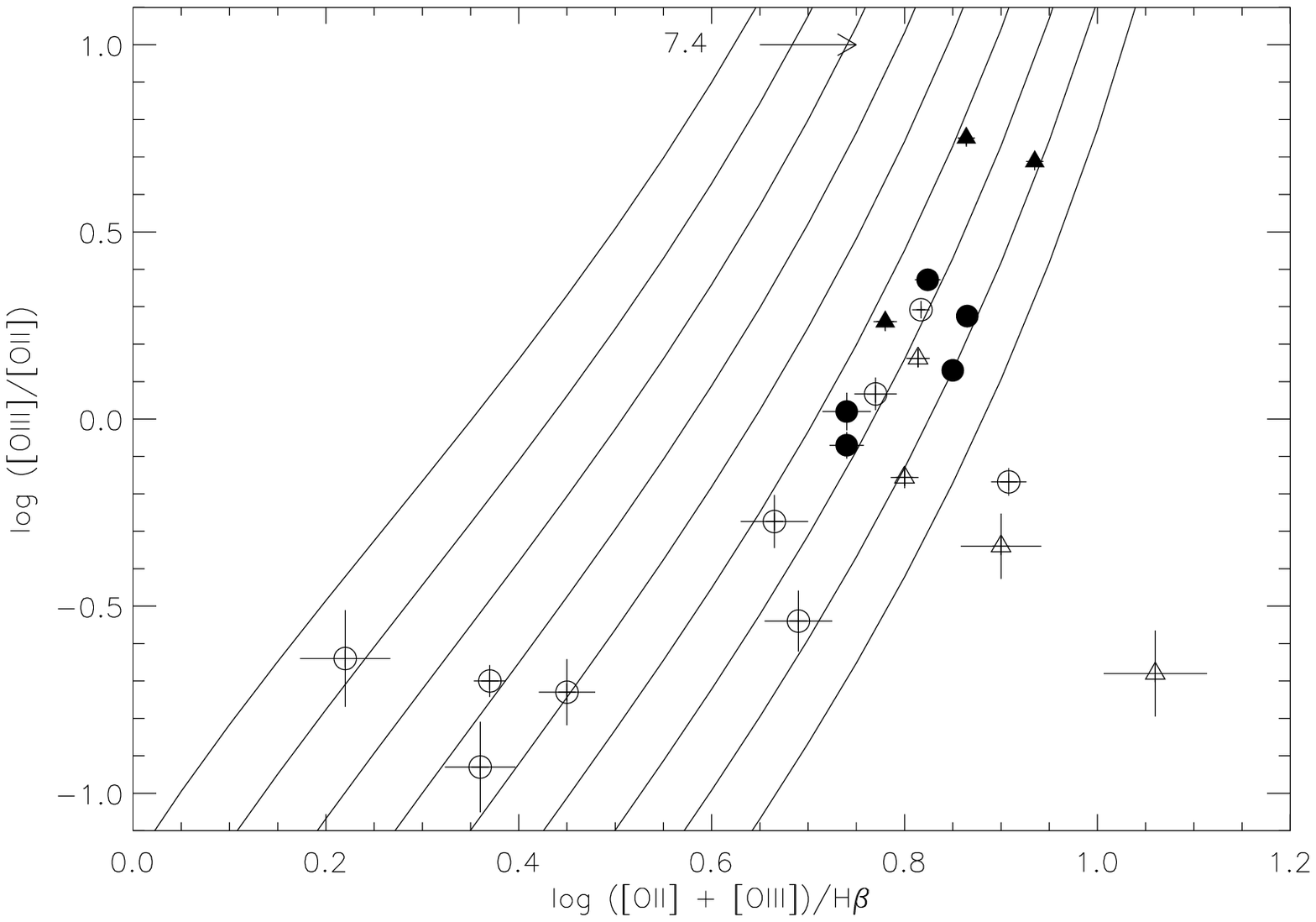}

\caption{log([O{\sc iii}]/[O{\sc ii}]) vs. logR$_{23}$ diagnostic diagram with McGaugh (1991) 
models, as reported in Kobulnicky et al. (1999), shown as continuous lines.
Each line correspond to an oxygen abundance, starting at 12 + log(O/H) = 7.4 
--marked for the first line to the left-- each line increasing to the right 
in the plot by 0.1 dex.
Points are coded as in Figure~3. Triangles represent sample objects corresponding to  
the low abundance models shown in this figure; while circles represent objects  
to be analyzed with models for higher abundances as shown in Figure~5 .}
\label{fig4} 
\end{figure}

\clearpage
\begin{figure}   
\epsscale{1.0}
\plotone{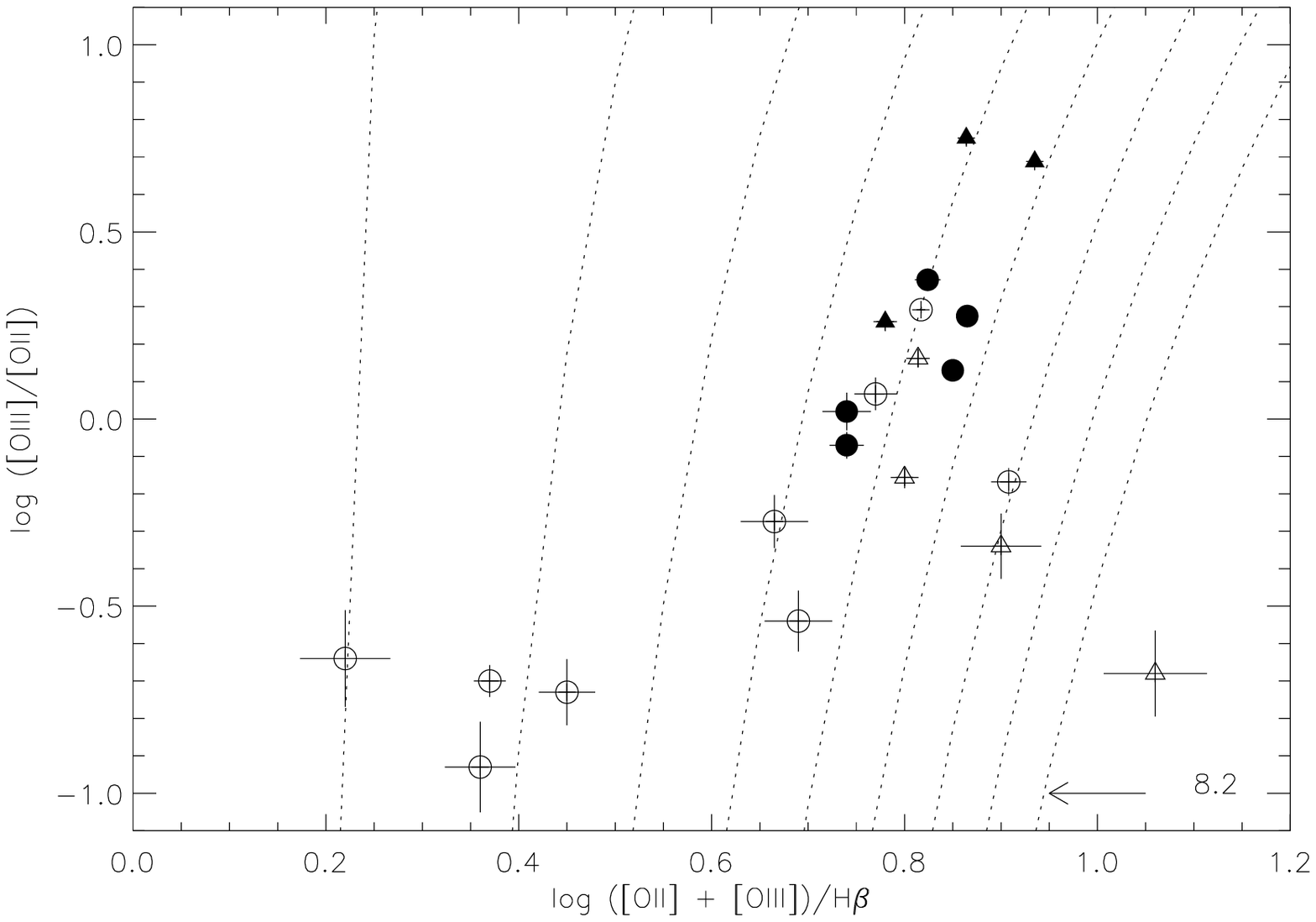}

\caption{log([O{\sc iii}]/[O{\sc ii}]) vs. log R$_{23}$ diagnostic diagram. McGaugh (1991) 
models, as reported in Kobulnicky et al. (1999), are shown as dotted lines. 
Each line correspond to an oxygen abundance, starting at 12 + log(O/H) = 8.2 
--marked for the first line to the right-- each line and increasing to the left in the 
plot by 0.1 dex.
Points are coded as in Figure~3. Circles represent sample galaxies corresponding to
the high abundance models shown in this figure. Triangles represent objects to be analyzed
with models for lower abundances as shown in Figure~4 .}
\label{fig5} 
\end{figure}

\clearpage

\begin{deluxetable}{lllcccrcr}
\tabletypesize{\footnotesize}
\tablecaption{Journal of the Observations\label{tab1}. (1) VCC name. 
(2) Other name. (3) Type (4),(5) 1950 coordinates. (6) Position Angle of the slit. (7) Central wavelength of the 
observation. (8) Exposure time. (9) Date of the observation.}
\tablehead{
  \colhead{VCC}    &
  \colhead{Other name}        &
  \colhead{Type}        &
  \colhead{$\alpha$ (1950)}    &
  \colhead{$\delta$ (1950)}        & 
  \colhead{Pos.Ang.} &
  \colhead{$\lambda_c$}  & 
  \colhead{Exposure}  & 
  \colhead{Date}     \\
  \colhead{}          &
  \colhead{}          &
  \colhead{($hh$ $mm$ $ss$)} &
  \colhead{($^\circ$ $\arcmin$ $\arcsec$)}  &   
  \colhead{($^\circ$)}  &   
  \colhead{(\AA )}    & 
  \colhead{(s)}       & 
  \colhead{}         
}
\startdata
1699 & IC~3591  & SBmIII & 12 34 30.3 & 07 12 10 & 237  & 4228 & 1400 & 1993 May 25 \\
         &&& & &        & 4228 & 1300 & 1993 May 26 \\
         &&& & &        & 6448 & 1300 & 1993 May 25 \\
         &&& & &        & 9100 & 1600 & 1993 May 26 \\
144 & Haro~6   &  BCD   & 12 12 45.1 & 06 02 24 &   310  & 4228 & 2400 & 1993 May 25 \\
         &&& & &        & 6448 & 2300 & 1993 May 25 \\
         &&& & &        & 4223 & 1800 & 1993 May 26 \\
         &&& & &        & 9096 & 1800 & 1993 May 26 \\
1313 & RMB~132  &BCD& 12 28 16.6 & 12 19 20 &   266  & 4228 & 2400 & 1993 May 25 \\
         &&& & &        & 6448 & 2400 & 1993 May 25 \\
562 & RMB~175  &BCD& 12 20 03.3 & 12 26 05 & 138    & 4228 & 3400 & 1993 May 24 \\
         &&& & &        & 6448 & 3000 & 1993 May 24 \\
2033 & A~1243+08&BCD& 12 43 32.8 & 08 45 02 &   237  & 4228 & 1183 & 1993 May 25 \\
         &&& & &        & 4228 & 1400 & 1993 May 26 \\
         &&& & &        & 6448 & 1162 & 1993 May 25 \\
324 & Mkn~49   &Epec/BCD& 12 16 36.6 & 04 07 59 &   3    & 4228 & 2200 & 1993 May 24 \\
         &&& & &        & 6448 & 2100 & 1993 May 24 \\
848 & A~1223+06&ImIIIpec/BCD & 12 23 19.5 & 06 05 11 &  24 & 4223 & 2400 & 1993 May 26 \\
         &&& & &                             & 6445 & 2300 & 1993 May 26 \\
         &&    & 12 23 19.6 & 06 05 09 &     & 4382 & 1700 & 1994 March 11 \\
         &&    &            &          &     & 6549 & 1600 & 1994 March 11  \\
1437 & A~1230+09&BCD& 12 30 01.5 & 09 26 58 &   57   & 4223 & 1800 & 1993 May 26 \\
         &&& & &        & 6445 & 1700 & 1993 May 26 \\
841 & RMB~46   &BCD& 12 23 15.9 & 15 13 45 &   344  & 4223 & 1800 & 1993 May 26 \\
         &&& & &        & 6445 & 1700 & 1993 May 26 \\
334 & RMB~56   &BCD& 12 16 41.7 & 12 26 05 &   3    & 4228 & 2000 & 1993 May 24 \\
         &&& & &        & 6448 & 1900 & 1993 May 24 \\
1374 & IC~3453  &IBm/BCD& 12 31 37.8 & 14 51 38 & 345 & 4382 & 1400 & 1994 March 10 \\
         &&    &            &          &     & 6549 & 1300 & 1994 March 10 \\
655 & NGC~4344 &SpN/BCD & 12 23 37.5 & 17 32 27 &  69 & 4382 & 1600 & 1994 March 10 \\
         && amorph.   &   &          &     & 6549 & 1500 & 1994 March 10 \\ 
1725 & A1235+08 &SmIII/BCD& 12 37 41.2 & 08 33 33 &116.5& 4382 & 1400 & 1994 March 12 \\
         &  &  &            &          &     & 6549 & 1300 & 1994 March 12 \\
135 & IC~3063  &S pec/BCD & 12 15 06.7 & 12 01 01 &20.5 & 4382 & 1243 & 1994 March 12 \\
         &&    &            &          &     & 6549 & 1229 & 1994 March 12 \\
1486 & IC~3483  &S pec?,N    & 12 30 38.1 & 11 37 22 &170.5& 4382 & 1400 & 1994 March 12 \\
         &&    &            &          &     & 6549 & 1300 & 1994 March 12 \\
1955 & NGC~4641 &Spec/BCD   & 12 43 97.6 & 12 03 03 &  28 & 4382 & 1400 & 1994 March 12 \\
         & & amorph.  &     &          &     & 6549 & 1300 & 1994 March 12 \\
2037 & 10$^\circ$~71 &ImIII/BCD& 12 46 15.3 & 10 12 12 & 156   & 4382 & 1800 & 1994 March 11 \\
              &&    &            &          &       & 6549 & 1700 & 1994 March 11 \\
72 & 15$^\circ$~9  &ImIII/BCD & 12 13 02.0 & 14 55 58 & 134.5 & 4382 & 1800 & 1994 March 10 \\
         &     &    &            &          &       & 6549 & 1700 & 1994 March 10 \\
428 & BB~18    &BCD & 12 20 40.2 & 13 53 20 &  29 & 4382 & 1500 & 1994 March 11 \\
         & &   &            &          &     & 6549 & 1500 & 1994 March 11 \\
802 & BO~146   &BCD & 12 25 28.7 & 13 29 50 &61.9 & 4382 & 1800 & 1994 March 11 \\
         &  &  &            &          &     & 6549 & 1700 & 1994 March 11 \\
213 & IC~3094  &dS?/BCD? & 12 16 56.0 & 13 37 33 &  90 & 4382 & 1600 & 1994 March 11 \\
     &         & amorph. &            &          &     & 6549 & 1500 & 1994 March 11 \\
1179 & IC~3412  &ImIII/BCD & 12 29 22.6 & 09 59 17 &  18 & 4382 & 1700 & 1994 March 11 \\
     &          &     &            &          &     & 6549 & 1600 & 1994 March 11 \\
\enddata
\end{deluxetable}

\clearpage

\begin{deluxetable}{lcrcccccc}
\tabletypesize{\footnotesize}
\tablecaption{Reddening corrected line intensities relative to H$\beta$ for the sample of Virgo 
dwarf galaxies. The reddening coefficient, $C$(H$\beta$), the equivalent width of H$\beta$, 
$EW$(H$\beta$) (\AA ), and the measured H$\beta$ flux, $F$(H$\beta$) 
in units 10$^{-15}$ erg cm$^{-2}$ s$^{-1}$, are included. Balmer lines were corrected from 
underlying stellar absorption. Errors quoted in parenthesis. 
\label{tab2}
}
\tablehead{
\colhead{Line } & 
\colhead{$\lambda$(\AA)} & 
\colhead{$f(\lambda)$} & 
\colhead{VCC~1699} & 
\colhead{VCC~144} & 
\colhead{VCC~1313} & 
\colhead{VCC~562} &
\colhead{VCC~2033}&
\colhead{VCC~324}
}
\startdata
[O {\sc ii}] 	        & 3727 & 0.26    & 1.260 & 2.541 & 1.102 & 2.656 & 3.016 & 2.215 \\
                        &      &         &(0.018)&(0.031)&(0.032)&(0.076)&(0.232)&(0.024)\\
H12 		        & 3751 & 0.20    & 0.019 & ---   & ---   & ---   & ---   & ---   \\
                        &      &         &(0.001)&       &       &       &       &       \\
H11 			& 3770 & 0.26    & 0.033 & ---   & ---   & ---   & ---   & ---   \\
                        &      &         &(0.002)&       &       &       &       &       \\
H10 			& 3798 & 0.25    & 0.042 & ---   & ---   & ---   & ---   & ---   \\
                        &      &         &(0.002)&       &       &       &       &       \\
H9 			& 3835 & 0.24    & 0.067 & ---   & ---   & ---   & ---   & ---   \\
                        &      &         &(0.003)&       &       &       &       &       \\

[Ne {\sc iii}] 		& 3868 & 0.23    & 0.364 & 0.270 & 0.338 & ---   & ---   & 0.183 \\
                        &      &         &(0.007)&(0.006)&(0.014)&       &       &(0.005)\\
H8  	                & 3889 & 0.22    & 0.177 & 0.154 & 0.167 & 0.100 & ---   & 0.109 \\
+He {\sc i}             &      &         &(0.004)&(0.005)&(0.010)&(0.012)&       &(0.003)\\

H$\epsilon$             & 3969 & 0.21 & 0.264 & 0.194 & ---   & ---   & ---   & 0.132 \\

+[Ne {\sc iii}]         &      &         &(0.005)&(0.005)&       &       &       &(0.004)\\
H$\delta$ 		& 4100 & 0.18    & 0.261 & 0.264 & 0.251 & 0.263 & 0.263 & 0.265 \\
                        &      &         &(0.005)&(0.006)&(0.010)&(0.015)&(0.048)&(0.005)\\
H$\gamma$ 		& 4340 & 0.14    & 0.471 & 0.469 & 0.469 & 0.466 & 0.470 & 0.469 \\
                        &      &         &(0.008)&(0.008)&(0.015)&(0.022)&(0.058)&(0.007)\\

[O {\sc iii}] 		& 4363 & 0.13    & 0.041 & 0.024 & 0.095 &$<0.008$&$<0.022$&$<0.001$\tablenotemark{u} \\
                        &      &         &(0.002)&(0.002)&(0.006)&       &       &       \\
He {\sc i} 		& 4471 & 0.10    & 0.036 & 0.028 & 0.027 & ---   & ---   & 0.029 \\
                        &      &         &(0.002)&(0.002)&(0.003)&       &       &(0.002)\\

[Ar {\sc iv}] 		& 4713 & 0.04    & 0.005 & ---   & ---   & ---   & ---   & ---   \\
                        &      &         &(0.001)&       &       &       &       &       \\
H$\beta$ 		& 4861 & 0.00    & 1.000 & 1.000 & 1.000 & 1.000 & 1.000 & 1.000 \\
                        &      &         &(0.013)&(0.013)&(0.024)&(0.038)&(0.086)&(0.012)\\
He {\sc i} 		& 4922 & $-0.01$ & 0.007 & ---   & ---   & ---   & ---   & ---   \\
                        &      &         &(0.001)&       &       &       &       &       \\

[O {\sc iii}] & 4959 & $-0.02$ & 1.819 & 1.177 & 1.570 & 0.948 & 0.345\tablenotemark{c}& 1.067\\
                        &      &         &(0.022)&(0.015)&(0.034)&(0.036)&       &(0.013)\\

[O {\sc iii}] 		& 5007 & $-0.03$ & 5.531 & 3.605 & 4.637 & 2.906 & 1.260 & 3.272 \\
                        &      &         &(0.052)&(0.037)&(0.081)&(0.079)&(0.109)&(0.032)\\
He {\sc i} 		& 5876 & $-0.23$ & 0.113 & 0.098 & 0.085 & ---   & ---   & 0.103 \\
                        &      &         &(0.003)&(0.003)&(0.006)&       &       &(0.003)\\

[O {\sc i}] 		& 6300 & $-0.30$ & 0.016 & 0.054 & 0.017 & ---   & ---   & 0.031 \\ 
                        &      &         &(0.002)&(0.002)&(0.003)&       &       &(0.002)\\

[S {\sc iii}] 		& 6312 & $-0.30$ & 0.016 & 0.012 & ---   & ---   & ---   & ---   \\
                        &      &         &(0.001)&(0.001)&       &       &       &       \\

[N {\sc ii}]            & 6548 & $-0.34$ & 0.025 & 0.069 & ---   & 0.025\tablenotemark{u} & ---  
& 0.061 \\
                        &      &         &(0.001)&(0.003)&       &       &       &(0.002)\\
H$\alpha$ 		& 6563 & $-0.34$ & 2.868 & 2.859 & 2.861 & 2.848 & 2.865 & 2.860 \\
                        &      &         &(0.031)&(0.030)&(0.051)&(0.080)&(0.190)&(0.030) \\

[N {\sc ii}]            & 6584 & $-0.34$ & 0.079 & 0.205 & 0.031 & 0.146\tablenotemark{c} & 
0.259 & 0.198 \\
                        &      &         &(0.003)&(0.003)&(0.003)&(0.012)&(0.039)&(0.004) \\
He {\sc i} 		& 6678 & $-0.35$ & 0.033 & 0.030 & 0.023 & ---   & ---   & 0.029 \\
                        &      &         &(0.002)&(0.001)&(0.002)&       &       &(0.002) \\

[S {\sc ii}] 		& 6717 & $-0.36$ & 0.104 & 0.181 & 0.116 & 0.298 & 0.859 & 0.244 \\
                        &      &         &(0.002)&(0.004)&(0.005)&(0.017)&(0.072)&(0.005) \\

[S {\sc ii}] 		& 6731 & $-0.36$ & 0.077 & 0.211 & 0.085 & 0.196 & 0.320 & 0.184 \\
                        &      &         &(0.003)&(0.004)&(0.004)&(0.013)&(0.039)&(0.004) \\
He {\sc i} 		& 7065 & $-0.40$ & 0.026 & 0.019 & 0.023 & ---   & ---   & ---   \\
                        &      &         &(0.001)&(0.002)&(0.003)&       &       &       \\

[Ar {\sc iii}] 		& 7135 & $-0.41$ & 0.105 & ---   & 0.046 & 0.007\tablenotemark{u}& ---   & 0.078 \\
                        &      &         &(0.003)&       &(0.004)&       &       &(0.003) \\
P14 			& 8598 & $-0.62$ & 0.005 & ---   & ---   & ---   & ---   & ---   \\
                        &      &         &(0.001)&       &       &       &       &       \\
P13 			& 8665 & $-0.63$ & 0.005\tablenotemark{u}& ---   & ---   & ---   & ---   & ---   \\
                        &      &         &       &       &       &       &       &       \\
P10 			& 9014 & $-0.67$ & 0.013 & ---   & ---   & ---   & ---   & ---   \\
                        &      &         &(0.002)&       &       &       &       &       \\

[S {\sc iii}] 		& 9069 & $-0.68$ & 0.157 & 0.101 & ---   & ---   & ---   & ---   \\
                        &      &         &(0.006)&(0.006)&       &       &       &       \\
P9 			& 9229 & $-0.70$ & 0.016\tablenotemark{u} & ---   & ---   & ---   & ---   & ---   \\
                        &      &         &       &       &       &       &       &       \\

[S {\sc iii}]		& 9532 & $-0.75$ & 0.550\tablenotemark{u}& 0.300 & ---   & ---   & ---   & ---   \\
                        &      &         &       &(0.014)&       &       &       &       \\
P8 			& 9546 & $-0.75$ & 0.025\tablenotemark{u}& ---   & ---   & ---   & ---   & ---   \\
                        &      &         &       &       &       &       &       &       \\

$C$(H$\beta$)           &      &         & 0.14  & 0.08  & 0.01  & 0.11  & 0.10  & 0.23 \\
$F$(H$\beta$)          &&      &54.4   &42.8   &12.1   &16.4   &2.14   &80.5\\
$EW$(H$\beta$)         &&      & 81    & 50    & 72    & 59    & 8     & 59 \\
\enddata
\tablenotetext{u}{Uncertain value}
\tablenotetext{c}{Cosmic ray}
\end{deluxetable}

\addtocounter{table}{-1}

\newpage


\begin{deluxetable}{lcrccccccc}
\tabletypesize{\footnotesize}
\tablecaption{Continued. 
}
\tablehead{
\colhead{Line } & 
\colhead{$\lambda$(\AA)} & 
\colhead{$f(\lambda)$} & 
\colhead{VCC~848\tablenotemark{1} a} & 
\colhead{ b } & 
\colhead{ c } & 
\colhead{VCC~848\tablenotemark{2}} & 
\colhead{VCC~1437} &
\colhead{VCC~841} &
\colhead{VCC~334}
}
\startdata
[O {\sc ii}] & 3727 & 0.26             & 1.989 & 2.968 & 2.573 & 3.410 & 4.816 & 2.714 & 3.725 \\ 
             &      &                  &(0.074)&(0.460)&(0.276)&(0.111)&(0.187)&(0.133)&(0.123)\\

[Ne {\sc iii}] & 3868 & 0.23           & 0.233 & ---   & ---   & ---   & ---   & ---   & ---   \\
             &      &                  &(0.034)&       &       &       &       &       &       \\
H8           & 3889 & 0.22             & 0.115 & ---   & ---   & ---   & ---   & ---   & --- \\
+ He {\sc i} &      &                  &(0.031)&       &       &       &       &       &       \\

H$\epsilon$  &3969  & 0.21             & 0.160 & ---   & ---   & ---   & ---   & ---   & --- \\

+[Ne {\sc iii}]            &      &    &(0.032)&       &       &       &       &       &       \\
H$\delta$ & 4100 & 0.18                & 0.261 & 0.262 & ---   &0.247\tablenotemark{u}& 0.263 & 0.264 & 0.181 \\
          &      &                     &(0.034)&(0.157)&       &(0.027)&(0.039)&(0.027)& (0.012) \\
H$\gamma$ & 4340 & 0.14                &0.467  &0.468  &0.470\tablenotemark{u}&0.472 & 0.466 & 0.466 & 0.473 \\
           &      &                    &(0.032)&(0.173)&       &(0.025)&(0.036)&(0.038)& (0.025) \\

[O {\sc iii}] & 4363 & 0.13          & 0.046 &$<0.079$&$<0.040$&0.041  & $<0.014$&$<0.014$&$<0.018$\\
           &      &                    &(0.015)&       &       &(0.014)&       &       &       \\
H$\beta$ & 4861 & 0.00                 & 1.000 & 1.000 & 1.000 & 1.000 &1.000 & 1.000 & 1.000 \\
           &      &                    &(0.044)&(0.240)&(0.124)&(0.038)&(0.060)&(0.062)& (0.040) \\

[O {\sc iii}] & 4959 & $-0.02$         &1.158  & 0.416 & 0.580 & 1.072 &0.791 & 0.722 & 0.596\tablenotemark{u}\\
           &      &                    &(0.049)&(0.168)&(0.094)&(0.041)&(0.053)&(0.051)&         \\

[O {\sc iii}] & 5007 & $-0.03$         & 3.526 & 1.524 & 1.180 & 3.211 & 2.482 & 2.448 & 2.005 \\
           &      &                    &(0.113)&(0.307)&(0.138)&(0.093)&(0.115)&(0.126)& (0.066) \\
He {\sc i} & 5876 & $-0.23$            & 0.060 & ---   & ---   & 0.104 & 0.095 & 0.084 & --- \\
           &      &                    &(0.012)&       &       &(0.018)&(0.016)&(0.016)&       \\

[O {\sc i}] & 6300 & $-0.30$           & 0.030 & ---   & ---   & 0.082 & ---   & 0.024 & --- \\ 
           &      &                    &(0.009)&       &       &(0.017)&       &(0.008)&       \\

[N {\sc ii}] & 6548 & $-0.34$  & 0.091\tablenotemark{u}& 0.111 & ---   &0.085\tablenotemark{u}& 0.145 & 0.102\tablenotemark{u} & 0.040\tablenotemark{u}\\
           &      &                    &       &(0.045)&       &(0.014)&(0.018)&       &    \\
H$\alpha$ & 6563 & $-0.34$             & 2.853 & 2.853 & 2.864 & 2.849 & 2.851 & 2.850 & 2.873 \\
           &      &                    &(0.090)&(0.430)&(0.243)&(0.098)&(0.152)&(0.158)& (0.090) \\

[N {\sc ii}] & 6584 & $-0.34$          & 0.177 & 0.369 & 0.373 & 0.196 & 0.515 & 0.370 & 0.223 \\
           &      &                    &(0.012)&(0.082)&(0.061)&(0.018)&(0.038)&(0.034)& (0.019) \\
He {\sc i} & 6678 & $-0.35$            & ---   & ---   & ---   & ---   &---   & 0.017 & --- \\
           &      &                    &       &       &       &       &      &(0.006)&       \\

[S {\sc ii}] & 6717 & $-0.36$          & 0.136 & 0.591 & 0.899 & 0.361 & 0.361 & 0.401 & 0.504 \\
           &      &                    &(0.012)&(0.114)&(0.100)&(0.022)&(0.031)&(0.035)& (0.026) \\

[S {\sc ii}] & 6731 & $-0.36$          & 0.107 & 0.448 & 0.474 & 0.253 & 0.276 & 0.280 & 0.358 \\
           &      &                    &(0.011)&(0.094)&(0.069)&(0.018)&(0.026)&(0.028)& (0.020) \\

[Ar {\sc iii}] & 7135 & $-0.41$        & ---   & ---   & ---   & 0.085 & 0.074 & 0.088 & --- \\
           &      &                    &       &       &       &(0.013)&(0.013)&(0.013)&       \\

$C$(H$\beta$)                        & &  & 0.07    & 0.12 & 0.06 & 0.22 & 0.43    & 0.40    & 0.13 \\
$F$(H$\beta$)                        & &  &6.71     &2.11  &1.55  & 14.2 & 12.4    &7.76     &8.31\\
$EW$(H$\beta$)                       & &  & 32      & 5    & 14   & 14   & 5       & 21      & 13 \\
\enddata
\tablenotetext{u}{Uncertain value}
\tablenotetext{1}{Spectra of knots a, b and c (figure~1); 1993 run}
\tablenotetext{2}{Integrated spectrum; 1994 run}
\end{deluxetable}

\addtocounter{table}{-1}

\newpage


\begin{deluxetable}{lcrcccccc}
\tabletypesize{\footnotesize}
\tablecaption{Continued. 
}
\tablehead{
\colhead{Line } & 
\colhead{$\lambda$(\AA)} & 
\colhead{$f(\lambda)$} & 
\colhead{VCC~1374} & 
\colhead{VCC~655} & 
\colhead{VCC~1725} & 
\colhead{VCC~135} &
\colhead{VCC~1486} &
\colhead{VCC~1955}
}
\startdata
[O {\sc ii}]  & 3727 & 0.26                  & 3.007 & 2.035 & 2.959 & 1.344 & 3.763 & 1.973 \\ 
           &      &                          &(0.065)&(0.140)&(0.129)&(0.113)&(0.264)&(0.064)\\
H$\delta$ & 4100 & 0.18                      & 0.266 & 0.266 & 0.261\tablenotemark{u} & 0.263\tablenotemark{u} & 0.263\tablenotemark{u} & 0.261\tablenotemark{u} \\
          &      &                           &(0.014)&(0.044)&(0.022) &(0.060) &(0.055) &(0.020) \\
H$\gamma$ & 4340 & 0.14                      & 0.470 &  ---  & 0.469  & 0.469  & 0.470  & 0.469 \\
          &      &                           &(0.017)&       &(0.029) &(0.051) &(0.061) &(0.022)\\

[O {\sc iii}] & 4363 & 0.13                  & 0.015 &$<0.041$& 0.035 &$<0.032$&$<0.030$&$<0.010$\\
              &      &                       & (0.006)&      &(0.012) &        &        & \\
H$\beta$ & 4861 & 0.00                       & 1.000 & 1.000 & 1.000  & 1.000  & 1.000  & 1.000 \\
         &      &                            &(0.026)&(0.075)& (0.046)& (0.073)& (0.103)&(0.036)\\

[O {\sc iii}] & 4959 & $-0.02$               & 1.036 &  ---  & 0.647  & 0.109: &  ---   & 0.105 \\
            &      &                         &(0.027)&       &(0.037) &        &        &(0.014)\\

[O {\sc iii}] & 5007 & $-0.03$               & 2.995 & 0.186 & 1.853  & 0.226  & 0.814  & 0.290 \\
            &      &                         &(0.059)&(0.039)&(0.070) &(0.048) &(0.095) &(0.019)\\
He {\sc i} & 5876 & $-0.23$                  & 0.111 & ---   & ---    &  ---   & ---    & 0.084 \\
           &      &                          &(0.012)&       &        &        &        &(0.018)\\

[O {\sc i}] & 6300 & $-0.30$                 &  ---  & ---   &  ---   &  ---   & ---    & 0.065 \\ 
           &      &                          &       &       &        &        &        &(0.013)\\

[N {\sc ii}] & 6548 & $-0.34$                & 0.077 & 0.226\tablenotemark{u} & 0.157\tablenotemark{u} & 0.258 & 0.238 & 0.208 \\
           &      &                          &(0.009)&(0.040)&(0.023) &(0.074) & (0.060)&(0.019)\\
H$\alpha$ & 6563 & $-0.34$                   & 2.880 & 2.871 & 2.867  &  2.866 &  2.869 & 2.866 \\
          &      &                           &(0.065)&(0.203)&(0.099) &(0.211) &(0.264) &(0.102)\\

[N {\sc ii}] & 6584 & $-0.34$                & 0.242 & 0.821 &  0.386 &  0.907 &  0.595 & 0.647 \\
           &      &                          &(0.013)&(0.078)& (0.030)& (0.084)& (0.089)&(0.034)\\

[S {\sc ii}] & 6717 & $-0.36$                & 0.413 & 0.548 &  0.572 &  0.466 &  0.690 & 0.468 \\
           &      &                          &(0.017)&(0.060)&(0.032) & (0.073)& (0.099)&(0.027)\\

[S {\sc ii}] & 6731 & $-0.36$                & 0.294 & 0.404 &  0.403 &  0.321 &  0.561 & 0.334 \\
           &      &                          &(0.015)&(0.051)& (0.027)& (0.063)& (0.091)&(0.022)\\

[Ar {\sc iii}] & 7135 & $-0.41$              & 0.087 & 0.068 & 0.117  &   ---  &  ---   & 0.046 \\
             &      &                        &(0.012)&(0.026)&(0.017) &        &        &(0.012)\\
$C$(H$\beta$)                    &     &     &  0.22 &  0.44 &  0.04  &  0.52  &  0.36  & 0.39 \\
$F$(H$\beta$)                        & &     & 23.2  &  9.17 &  7.34  &  5.85  &  3.67  &14.8\\
$EW$(H$\beta$)                       & &     & 16    &  4    & 18     &  4     &  5     &12 \\
\enddata
\tablenotetext{u}{Uncertain value}
\end{deluxetable}

\addtocounter{table}{-1}

\newpage


\begin{deluxetable}{lcrcccccc}
\tabletypesize{\footnotesize}
\tablecaption{Continued. 
}
\tablehead{
\colhead{Line } & 
\colhead{$\lambda$(\AA)} & 
\colhead{$f(\lambda)$} & 
\colhead{VCC~2037} & 
\colhead{VCC~72} & 
\colhead{VCC~428} &
\colhead{VCC~802} &
\colhead{VCC~213} &
\colhead{VCC~1179}
}
\startdata
[O {\sc ii}] & 3727 & 0.26                & 5.438 & 3.424 & 2.128 & 2.696 & 2.361 & 9.503 \\ 
             &      &                     &(0.488)&(0.408)&(0.070)&(0.160)&(0.121)&(1.124)\\

[Ne {\sc iii}] & 3868 & 0.23              &  ---  & ---   & 0.214 & 0.303 & ---   & --- \\
             &      &                     &       &       &(0.023)&(0.053)&       & \\

H8 + He {\sc i} & 3889 & 0.22             &  ---  & ---   & 0.150 & 0.203 & ---   & --- \\
                &      &                  &       &       &(0.022)&(0.048)&       & \\
H$\epsilon$     & 3969 & 0.21             &  ---  &  ---  & 0.188 &  ---  & ---   & --- \\

+ [Ne {\sc iii}]&      &                 &        &       &(0.022)&       &       & \\
H$\delta$ & 4100 & 0.18                   & ---   & 0.261\tablenotemark{u}& 0.262 & 0.259\tablenotemark{u}& 0.261 & 0.262 \\
          &      &                        &       &(0.109)&(0.017)&(0.036)&(0.044)&(0.116)\\
H$\gamma$ & 4340 & 0.14                   & 0.469 & 0.467 & 0.470 & 0.468 &  ---  & 0.469\tablenotemark{u} \\
          &      &                        &(0.095)&(0.126)&(0.022)&(0.041)&       &(0.126) \\

[O {\sc iii}] & 4363 & 0.13              &$<0.047$&$<0.063$& 0.072 & 0.031 &$<0.028$&$<0.055$\\
            &      &                      &       &       &(0.012)&(0.013)&       & \\
H$\beta$ & 4861 & 0.00                    & 1.000 & 1.000 & 1.000 & 1.000 & 1.000 & 1.000 \\
         &      &                         &(0.128)&(0.149)&(0.031)&(0.068)&(0.065)&(0.169)\\

[O {\sc iii}] & 4959 & $-0.02$            & 0.682 &  ---  & 0.963 & 0.700 & 0.109 & 0.471 \\
              &      &                    &(0.111)&       &(0.031)&(0.055)&(0.043)&(0.119)\\

[O {\sc iii}] & 5007 & $-0.03$            & 1.812 & 0.477 & 2.910 & 2.150 & 0.336 & 1.525 \\
            &      &                      &(0.200)&(0.105)&(0.072)&(0.121)&(0.051)&(0.223)\\
He {\sc i} & 5876 & $-0.23$               &  ---  &  ---  & ---   & ---   & 0.219 & --- \\
           &      &                       &       &       &       &       &(0.043)& \\

[O {\sc i}] & 6300 & $-0.30$              &  ---  & ---   &  ---  & ---   & ---   & --- \\ 
          &      &                        &       &       &       &       &       & \\

[N {\sc ii}] & 6548 & $-0.34$             &  ---  & 0.315\tablenotemark{u}& --- & 0.103 & 0.259 & 0.216\tablenotemark{u}\\
           &      &                       &       &(0.131)&       &(0.031)&(0.039)&(0.096)\\
H$\alpha$ & 6563 & $-0.34$                & 2.863 & 2.861 & 2.871 & 2.858 & 2.858 & 2.864 \\
          &      &                        &(0.345)&(0.363)&(0.083)&(0.179)&(0.180)&(0.363)\\

[N {\sc ii}] & 6584 & $-0.34$             & 0.354 & 0.729 & 0.072 & 0.355 & 0.910 & 0.434 \\
           &      &                       &(0.109)&(0.170)&(0.012)&(0.046)&(0.076)&(0.117)\\

[S {\sc ii}] & 6717 & $-0.36$             & 0.781 & 1.525\tablenotemark{u} & 0.173 & 0.481 & 0.473 & 0.710 \\
           &      &                       &(0.129)&(0.243)&(0.015)&(0.053)&(0.052)&(0.146)\\

[S {\sc ii}] & 6731 & $-0.36$             & 0.630 & 0.445 & 0.135 & 0.355 & 0.364 & 0.559 \\
           &      &                       &(0.113)&(0.143)&(0.014)&(0.046)&(0.046)&(0.130)\\
He {\sc i} & 7065 & $-0.40$               & ---   & ---   & 0.037 & ---   &  ---  & --- \\
           &      &                       &       &       &(0.009)&       &       & \\

[Ar {\sc iii}] & 7135 & $-0.41$           & ---   &  ---  & 0.061 &  ---  & ---   & 0.209 \\
             &      &                     &       &       &(0.010)&       &       &(0.099)\\
$C$(H$\beta$) &     &                     &  0.30 &  0.10 &  0.23 &  0.25 &  0.48 & 0.04 \\
$F$(H$\beta$)                        &&   &  1.54 & 0.755 & 12.5  &  3.13 & 14.3  & 0.655 \\
$EW$(H$\beta$)                       &&   &  6    &  ---  & 63    & 21    &  5    & 2 \\
\enddata
\hrule
\tablenotetext{u}{Uncertain value}
\end{deluxetable}

\clearpage

\begin{deluxetable}{rccl}
\tablecaption{Comparison of values of log R$_{23}$ with previous works. (1) Galaxy VCC number; 
(2) IG89: Izotov and Guseva (1989); (3) GH89: Gallagher and Hunter (1989); 
(4) this work. \label{tab3}}
\tablehead{
  \colhead{ VCC } &
  \colhead{ IG89 } &
  \colhead{ GH89 }    & 
  \colhead{this work}    
}
\startdata
144  &0.86 &0.89 & 0.865 \\
324  &0.72 &0.88 & 0.817 \\
334  & --  &0.91 & 0.80 \\
848  & --  &0.73\tablenotemark{u} & 0.824 \\
1374 & --  &0.92\tablenotemark{u} & 0.85 \\
1437 &0.74 & --  & 0.908 \\
1725 & --  &0.81 & 0.74 \\
2033 &0.89 &0.91 & 0.665 \\
\enddata
\tablenotetext{u}{uncertain value}
\end{deluxetable}

\clearpage

\begin{deluxetable}{lccccccccc}
\rotate
\tabletypesize{\tiny}
\tablecaption{Direct determination of physical conditions and abundances\label{tab4}
}
\tablehead{
\colhead{  } & 
\colhead{VCC~1699} & 
\colhead{VCC~144} & 
\colhead{VCC~1313} & 
\colhead{VCC~848\tablenotemark{a}}& 
\colhead{VCC~848\tablenotemark{i}}&
\colhead{VCC~1374} &
\colhead{VCC~1725} &
\colhead{VCC~428} &
\colhead{VCC~802}
}
\startdata
Te$_{[O{\sc III}]}$ (10$^4$ K)
&1.048$\pm$0.021&1.014$\pm$0.037&1.542$\pm$0.038&1.278$\pm$0.132&1.265$\pm$0.131&0.929$\pm$0.090&1.481$\pm$0.150
&1.696$\pm$0.10&1.33$\pm$0.14\\

Te$_{[S{\sc III}]}$ (10$^4$ K)
    &1.06$\pm$0.03& 1.15$\pm$0.03&      --    &  --        & -- & -- & -- &-- & \\
log Ne$_{[S{\sc II}]}$ (cm$^{-3}$)
   & 1.9        & 1.8        & 1.8        & 2.2      & $\le$2 & $\le$2 & $\le$2 & 2.1 & 1.8 \\
 &&&&&&&&& \\
12 + log (O$^+$/H$^+$)       
    & 7.61$\pm$0.05&8.00$\pm$0.07&7.08$\pm$0.07&7.57$\pm$0.15&7.72$\pm$0.16&7.94$\pm$0.16&7.54$\pm$0.15
&7.30$\pm$0.10&7.58$\pm$0.16 \\
12 + log (O$^{++}$/H$^+$)    
    &8.23$\pm$0.06&8.09$\pm$0.05&7.67$\pm$0.05&7.76$\pm$0.12&7.74$\pm$0.16&8.14$\pm$0.10&7.31$\pm$0.11
&7.36$\pm$0.08&7.50$\pm$0.14 \\
12 + log (O/H)               
    & 8.32$\pm$0.06&8.35$\pm$0.07&7.77$\pm$0.06&7.98$\pm$0.14&8.03$\pm$0.16&8.35$\pm$0.14&7.74$\pm$0.14
&7.64$\pm$0.09&7.84$\pm$0.15 \\
&&&&&&&&& \\
log (N$^+$/O$^+$)            
&-1.46$\pm$0.05&-1.39$\pm$0.04&-1.63$\pm$0.08&-1.22$\pm$0.08&-1.37$\pm$0.11&-1.33$\pm$0.09&-0.97$\pm$0.10
&-1.53$\pm$0.10&-1.0$\pm$0.12 \\
&&&&&&&&& \\
12 + log (S$^+$/H$^+$)       
    & 5.58$\pm$0.05&6.04$\pm$0.04&5.36$\pm$0.06&5.58$\pm$0.14&5.90$\pm$0.16&6.31$\pm$0.12&6.00$\pm$0.11&5.39$\pm$0.11&6.02$\pm$0.17  \\
12 + log (S$^{++}$/H$^+$)    
    & 6.57$\pm$0.08& 6.40$\pm$0.09 &   --      & -- & -- & -- & -- & -- & -- \\
12 + log (S/H)               
&$\ge$6.61$\pm$0.08&$\ge$6.56$\pm$0.08&  --    & -- & -- & -- & -- & -- & -- \\
&&&&&&&&& \\
log (Ne$^{++}$/O$^{++}$)    
 &-0.70$\pm$0.03&-0.64$\pm$0.06&-0.74$\pm$0.07&-0.74$\pm$0.08&     -- &  --  & --  &-0.74$\pm$0.10&-0.46$\pm$0.17 \\
&&&&&&&&& \\
log (Ar$^{++}$/O$^{++}$)     
 &-2.29$\pm$0.08&          --  &-2.42$\pm$0.09&     --      &-2.06$\pm$0.09&-2.16$\pm$0.10&-1.62$\pm$0.12&-2.06$\pm$0.10& -- \\
&&&&&&&&& \\
\enddata
\tablenotetext{a}{Knot a, 1993 run}
\tablenotetext{i}{Integrated spectrum, 1994 run}
\end{deluxetable}

\clearpage

\begin{deluxetable}{lcccccccccc}
\rotate
\tabletypesize{\scriptsize}
\tablecaption{Adopted abundances and ionization structure parameters for Virgo 
dwarf galaxies. \label{tab5}}
\tablehead{
  \colhead{ VCC } &
  \colhead{1699}    & 
  \colhead{ 144}    &
  \colhead{1313}        & 
  \colhead{ 562} & 
  \colhead{2033}    & 
  \colhead{ 324} & 
  \colhead{ 848}        & 
  \colhead{1437}  & 
  \colhead{ 841} &    
  \colhead{ 334} 
}
\startdata
 log R$_{23}$     &0.935&0.865& 0.864& 0.814& 0.665 & 0.817 & 0.824& 0.908 & 0.770 & 0.80 \\

 log ([O{\sc iii}]/[O{\sc ii}])&0.688 &0.275&0.751 & 0.162& -0.274& 0.292 & 0.372& -0.168& 0.067 & -0.156\\

 log ([N{\sc ii}]/[O{\sc ii}]) & -1.10 & -0.96 & -1.40 & -1.15\tablenotemark{u}& -0.96\tablenotemark{a} & 
-0.92 & -0.92 & -0.85 & -0.77 & -1.15 \\
&&&&&&&&&\\

12 + log (O/H) &&&&&&&&&\\

[OIII] $\lambda$4363 \tablenotemark{1}
  & 8.32& 8.35 & 7.77 &$\ge$8.5&$\ge$7.7& --  & 7.98 (8.03)\tablenotemark{i}&$\ge$8.5&$\ge$8.31& $\ge$8.1
\\  

P$_{upper}$\tablenotemark{2}
               & 8.38 & 8.35  & 8.49  & 8.38  & 8.40    & 8.42    &8.44 (8.25)\tablenotemark{i}& 8.10  & 8.40  & 8.27
\\

P$_{lower}$\tablenotemark{2}
               & 7.85  & 7.93  &  7.73 & 7.93  & 8.11   &  7.85   &7.81 (8.05)\tablenotemark{i}& 8.05  &  7.93 & 8.18
\\   

R$_{23}$$_{upper}$\tablenotemark{3} 
               & 8.52  & 8.54  & 8.61  & 8.58  & 8.71   & 8.60    &8.60 (8.5)\tablenotemark{i}& 8.41  & 8.63  & 8.56
\\  

R$_{23}$$_{lower}$\tablenotemark{3}
               & 8.09  & 8.08  & 7.92  & 8.02  & 7.93   & 7.99    &7.97 (7.9)\tablenotemark{i}& 8.30  & 7.98  & 8.11 
\\

[N{\sc ii}]/[O{\sc ii}]\tablenotemark{4} 
               & 8.5   & 8.5   & 7.9\tablenotemark{u}   & 8.5   & 8.6    & 8.6     & 8.5  &  8.6  & 8.7   & 8.5
\\

   S$_{23}$\tablenotemark{5} 
               & 8.20  & 8.10  &  --   &  --   & --     &  --     & --   &   --  &   --  & --
\\  
Adopted&&&&&& &&&& 
\\

$<$12 + log (O/H)$>$ &8.32  & 8.35  & 7.77 &8.6-8.4\tablenotemark{a}&8.7-8.4\tablenotemark{a}& 8.6-8.4& 8.0  &8.3$\pm$0.2&8.7-8.4&8.2-8.1
\\

$<$log (N/O)$>$    & -1.46 & -1.39 & -1.63&-1.6\tablenotemark{a} & -1.5\tablenotemark{a} &  -1.4   & -1.30 &  -1.3 &  -1.3 & -1.4
\\
\enddata

\tablenotetext{u}{Uncertain value}

\tablenotetext{i}{Abundances derived from the integrated spectrum of A1223+06 (in parenthesis)}

\tablenotetext{a}{At the 1$\sigma$ error level, [N{\sc ii}]/[O{\sc ii}] $\le$ 0.1 could be compatible, 
leading to $<$12 + log (O/H)$>$ = 8.0 , $<$log (N/O)$>$ = -1.4}

\tablenotetext{1}{Direct abundance determination (or lower limit) from the [OIII] $\lambda$4363 line flux}

\tablenotetext{2}{P-method calibration, upper-lower branch after Pilyugin (2000; 2001)}

\tablenotetext{3}{Calibration of R$_{23}$ (cf. Pagel et al. 1980), according to the theoretical models of 
McGaugh (1991): upper-lower branch as parameterized in Kobulnicky et al. (1999)}

\tablenotetext{4}{Calibration of the [N{\sc ii}]/[O{\sc ii}] abundance diagnostic 
according to the theoretical models of Dopita et al. (2000)} 

\tablenotetext{5}{Empirical calibration using sulphur lines, after D\' \i az and P\' erez-Montero (2000)}

\end{deluxetable}

\addtocounter{table}{-1}

\clearpage

\begin{deluxetable}{lccccccccccc}
\rotate
\tabletypesize{\scriptsize}
\tablecaption{ Continued. }
\tablehead{
  \colhead{ VCC } &
  \colhead{1374}    & 
  \colhead{ 655}    &
  \colhead{1725}     & 
  \colhead{ 135} & 
  \colhead{1486}    & 
  \colhead{1955} & 
  \colhead{2037}        & 
  \colhead{ 428} &
  \colhead{ 802}  & 
  \colhead{ 213} &    
  \colhead{1179}  
}
\startdata
log R$_{23}$    & 0.85 &0.36 & 0.74 & 0.22 & 0.69 & 0.37 & 0.90 & 0.78 & 0.74 & 0.45 & 1.06  \\

Log ([O{\sc iii}]/[O{\sc ii}])& 0.13 &-0.93&-0.07 &-0.64 &-0.54 &-0.70 &-0.34 & 0.26 & 0.02 &-0.73 & -0.68 \\

log ([N{\sc ii}]/[O{\sc ii}]) & -0.98 & -0.28 & -0.77 & -0.06 & -0.69 & -0.37 & -1.07 & 
-1.35 & -0.77 & -0.30 & -1.22 \\

&&&&&&&&&&&\\
12 + log (O/H) &&&&&&&&&&&\\

[OIII] $\lambda$4363 \tablenotemark{1}
  & 8.35 & --  & 7.74 & -- &$\ge$7.3  & --  & $\ge$7.7 & 7.64  & 7.84   & -- &$\ge$7.7 \\  

P$_{upper}$\tablenotemark{2}
               & 8.3   &   8.6 & 8.4   & 8.7   & 8.3    & 8.6     & 8.1  & 8.5  & 8.4   & 8.5   &  \tablenotemark{u}\\

P$_{lower}$\tablenotemark{2}
               & 8.0   &   8.5 & 8.0   & 7.9   & 8.4    & 8.2     & 8.5  & 7.8  & 7.9   & 8.4   & \tablenotemark{u} \\
   
R$_{23}$$_{upper}$\tablenotemark{3} 
               & 8.5   &  8.9  &  8.6  & 8.9   & 8.6    & 8.9     & 8.4  & 8.6  &  8.6  & 8.9   & 8.2  \\  

R$_{23}$$_{lower}$\tablenotemark{3}
               & 8.1   &  7.75 &  8.0  & 7.5   & 8.1    & 7.7     & 8.3  & 7.9  &  7.9  & 7.8   & 8.2  \\

[N{\sc ii}]/[O{\sc ii}]\tablenotemark{4} 
               & 8.6   & 8.8   & 8.7   & 8.9   & 8.7    & 8.9     & 8.4  & 8.1  &  8.7  & 8.9   & 8.0  \\
Adopted &  &&&&&&&&&& 
\\

$<$12 + log (O/H)$>$&8.35  &8.9-8.6 & 7.74 & 8.9-8.7 &8.7-8.3&8.9-8.6&8.3$\pm$0.2 & 7.64&  7.84 &8.9-8.5 &8.2$\pm$0.2  \\
 
$<$log (N/O)$>$   & -1.33 & -0.9  &-0.97 &-0.75& -1.2 &  -1.0   & -1.5 &-1.53& -1.0  &  -0.9 &  -1.5 \\

\enddata

\tablenotetext{u}{Uncertain value or calibration}

\tablenotetext{1}{Direct abundance determination (or lower limit) from the [OIII] $\lambda$4363 line flux}

\tablenotetext{2}{P-method calibration, upper-lower branch after Pilyugin (2000; 2001)}

\tablenotetext{3}{Calibration of R$_{23}$ (cf. Pagel et al. 1980), according to the theoretical models of 
McGaugh (1991): upper-lower branch as parameterized in Kobulnicky et al. (1999)}

\tablenotetext{4}{Calibration of the [N{\sc ii}]/[O{\sc ii}] abundance diagnostic 
according to the theoretical models of Dopita et al. (2000)} 

\tablenotetext{5}{Empirical calibration using sulphur lines, after D\' \i az and P\' erez-Montero (2000)}

\end{deluxetable}


\clearpage

\begin{thebibliography}{}
\bibitem[Allende Prieto, Lambert, \& Asplund(2001)]{2001ApJ...556L..63A} 
Allende Prieto, C., Lambert, D.~L., \& Asplund, M.\ 2001, \apjl, 556, L63 

\bibitem[Aller(1984)]{1984ptgn.conf.....A} Aller, L.~H.\ 1984, Physics of Thermal Gaseous Nebulae,
 1984 Astrophysics \& Space Science Library vol.~112

\bibitem[Alloin, Collin-Souffrin, Joly, \& 
Vigroux(1979)]{1979A&A....78..200A} Alloin, D., Collin-Souffrin, S., Joly, 
M., \& Vigroux, L.\ 1979, \aap, 78, 200 

\bibitem[Binggeli, Sandage, \& Tammann(1985)]{1985AJ.....90.1681B} Binggeli, B.,
Sandage, A., \& Tammann, G.~A.\ 1985, \aj, 90, 1681

\bibitem[Binggeli, Tammann, \& Sandage(1987)]{1987AJ.....94..251B} Binggeli, B.,
Tammann, G.~A., \& Sandage, A.\ 1987, \aj, 94, 251

\bibitem[Boselli et al.(1997)]{1997A&A...327..522B} Boselli, A., Gavazzi, G.,
Lequeux, J., Buat, V., Casoli, F., Dickey, J., \& Donas, J.\ 1997, \aap, 327,
522

\bibitem[Boselli et al. (2002)]{2002A&A...386..134B} Boselli, A., 
Iglesias-P{\' a}ramo, J., V{\' \i}lchez, J.~M.,\& Gavazzi, G.\ 2002, \aap, 386,
134

\bibitem[Bresolin \& Kennicutt(2002)]{2002ApJ...} 
Bresolin, F., \& Kennicutt, R.~C.\ 2002, \apj, in press astro-ph/0202383

\bibitem[Brosch, Almoznino, \& Hoffman(1998)]{1998A&A...331..873B} Brosch, N.,
Almoznino, E., \& Hoffman, G.~L.\ 1998, \aap, 331, 873

\bibitem[Charlot \& Longhetti(2001)]{2001MNRAS.323..887C} Charlot, S.~\& 
Longhetti, M.\ 2001, \mnras, 323, 887 

\bibitem[Castellanos, D{\' i}az, \& Terlevich(2002)]{2002MNRAS.329..315C} 
Castellanos, M., D{\' \i}az, A.~I., \& Terlevich, E.\ 2002, \mnras, 329, 315 

\bibitem[Cayatte, van Gorkom, Balkowski, \& Kotanyi(1990)]{1990AJ....100..604C}
Cayatte, V., van Gorkom, J.~H., Balkowski, C., \& Kotanyi, C.\ 1990, \aj, 100,
604

\bibitem[Contini, Treyer, Sullivan, \& Ellis(2002)]{2002MNRAS.330...75C} 
Contini, T., Treyer, M.~A., Sullivan, M., \& Ellis, R.~S.\ 2002, \mnras, 
330, 75 

\bibitem[D\' \i az \& P{\' e}rez-Montero(2000)]{2000MNRAS.312..130D} D\' \i az,
A.~I.~\& P{\' e}rez-Montero, E.\ 2000, \mnras, 312, 130

\bibitem[Dopita \& Evans(1986)]{1986ApJ...307..431D} Dopita, M.~A.~\& 
Evans, I.~N.\ 1986, \apj, 307, 431 

\bibitem[Dopita, Kewley, Heisler, \& Sutherland(2000)]{2000ApJ...542..224D}
Dopita, M.~A., Kewley, L.~J., Heisler, C.~A., \& Sutherland, R.~S.\ 2000, \apj,
542, 224

\bibitem[Duc \& Mirabel(1999)]{1999IAUS..186...61D} Duc, P.-A.~\& Mirabel,
I.~F.\ 1999, IAU Symp.~186: Galaxy Interactions at Low and High Redshift, 186,
61

\bibitem[Dutil \& Roy(1999)]{1999ApJ...516...62D} Dutil, Y.~\& Roy, J.~ 
1999, \apj, 516, 62 

\bibitem[Edmunds \& Pagel(1978)]{1978MNRAS.185P..77E} Edmunds, M.~G.~\& 
Pagel, B.~E.~J.\ 1978, \mnras, 185, 77P 

\bibitem[Edmunds \& Pagel(1984)]{1984MNRAS.211..507E} Edmunds, M.~G.~\& 
Pagel, B.~E.~J.\ 1984, \mnras, 211, 507 

\bibitem[Esteban, Peimbert, Torres-Peimbert, \& 
Escalante(1998)]{1998MNRAS.295..401E} Esteban, C., Peimbert, M., 
Torres-Peimbert, S., \& Escalante, V.\ 1998, \mnras, 295, 401 

\bibitem[Fioc \& Rocca-Volmerange(1997)]{1997A&A...326..950F} Fioc, M.~\& 
Rocca-Volmerange, B.\ 1997, \aap, 326, 950 

\bibitem[Gallagher \& Hunter(1984)]{1984ARA&A..22...37G} Gallagher, 
J.~S.~\& Hunter, D.~A.\ 1984, \araa, 22, 37. 

\bibitem[Gallagher \& Hunter(1989)]{1989AJ.....98..806G} Gallagher, J.~S.~\&
Hunter, D.~A.\ 1989, \aj, 98, 806

\bibitem[Garnett(1992)]{1992AJ....103.1330G} Garnett, D.~R.\ 1992, \aj, 103,
1330

\bibitem[Gavazzi et al.(2001)]{2001ApJ...563L..23G} Gavazzi, G., Boselli, A., Mayer, L., 
Iglesias-P\'{a}ramo, J., V{\' \i}lchez, J.~M., \& Carrasco, L.\ 2001, \apjl,
563, L23

\bibitem[Grogin \& Geller(2000)]{2000AJ....119...32G} Grogin, N.~A.~\& Geller,
M.~J.\ 2000, \aj, 119, 32

\bibitem[Hashimoto, Oemler, Lin, \& Tucker(1998)]{1998ApJ...499..589H}
Hashimoto, Y., Oemler, A.~J., Lin, H., \& Tucker, D.~L.\ 1998, \apj, 499, 589

\bibitem[Haynes, Giovanelli, \& Chincarini(1984)]{1984ARA&A..22..445H} Haynes,
M.~P., Giovanelli, R., \& Chincarini, G.~L.\ 1984, \araa, 22, 445

\bibitem[Henry, Edmunds, \& K{\" o}ppen(2000)]{2000ApJ...541..660H} Henry, 
R.~B.~C., Edmunds, M.~G., \& K{\" o}ppen, J.\ 2000, \apj, 541, 660 

\bibitem[Hidalgo-Gamez \& Olofsson(1998)]{1998A&A...334...45H}
Hidalgo-G\'{a}mez, A.~M.~\& Olofsson, K.\ 1998, \aap, 334, 45

\bibitem[Hoffman, Helou, \& Salpeter(1988)]{1988ApJ...324...75H} Hoffman, G.~L.,
Helou, G., \& Salpeter, E.~E.\ 1988, \apj, 324, 75

\bibitem[Holweger(2001)]{2001sgc..conf...23H} Holweger, H.\ 2001, Joint 
SOHO/ACE workshop "Solar and Galactic Composition", 23 

\bibitem[Hummer \& Storey(1987)]{1987MNRAS.224..801H} Hummer, D.~G.~\& Storey,
P.~J.\ 1987, \mnras, 224, 801

\bibitem[Iglesias-P{\' a}ramo \& V{\' i}lchez(1999)]{1999ApJ...518...94I}
Iglesias-P{\' a}ramo, J.~\& V{\' \i}lchez, J.~M.\ 1999, \apj, 518, 94

\bibitem[]{515} Iglesias-P\'{a}ramo, J., Boselli, A., Cortese, L., V\'{\i}lchez,
J.M., \& Gavazzi, G.\ 2002, \aap, 384, 383

\bibitem[Izotov \& Guseva(1989)]{1989Afz....30..564I} Izotov, Y.~I.~\& Guseva,
N.~G.\ 1989, Astrofizika, 30, 564

\bibitem[Kenney \& Young(1988)]{1988ApJS...66..261K} Kenney, J.~D.~\& Young,
J.~S.\ 1988, \apjs, 66, 261

\bibitem[King (19885)]{1985 LPTN} King,~D.~1985, La Palma Tech. Note, 15 

\bibitem[Kinkel \& Rosa(1994)]{1994A&A...282L..37K} Kinkel, U.~\& Rosa, 
M.~R.\ 1994, \aap, 282, L37 

\bibitem[Kobulnicky, Kennicutt, \& Pizagno(1999)]{1999ApJ...514..544K}
Kobulnicky, H.~A., Kennicutt, R.~C., \& Pizagno, J.~L.\ 1999, \apj, 514, 544

\bibitem[K{\" o}ppen \& Edmunds(1999)]{1999MNRAS.306..317K} K{\" o}ppen, 
J.~\& Edmunds, M.~G.\ 1999, \mnras, 306, 317 

\bibitem[Lee, Richer, \& McCall(2000)]{2000ApJ...530L..17L} Lee, H., Richer,
M.~G., \& McCall, M.~L.\ 2000, \apjl, 530, L17

\bibitem[Leitherer et al.(1999)]{1999ApJS..123....3L} Leitherer, C.~et al.\ 
1999, \apjs, 123, 3 

\bibitem[McCall(1984)]{1984MNRAS.208..253M} McCall, M.~L.\ 1984, \mnras, 208,
253
\bibitem[McCall, Rybski, \& Shields(1985)]{1985ApJS...57....1M} McCall, 
M.~L., Rybski, P.~M., \& Shields, G.~A.\ 1985, \apjs, 57, 1 

\bibitem[McGaugh(1991)]{1991ApJ...380..140M} McGaugh, S.~S.\ 1991, \apj, 380,
140

\bibitem[McGaugh(1994)]{1994ApJ...426..135M} McGaugh, S.~S.\ 1994, \apj, 
426, 135 

\bibitem[Miller \& Hodge(1996)]{1996ApJ...458..467M} Miller, B.~W.~\& 
Hodge, P.\ 1996, \apj, 458, 467 

\bibitem[Olofsson(1997)]{1997A&A...321...29O} Olofsson, K.\ 1997, \aap, 
321, 29 

\bibitem[Pagel et al.(1979)]{1979MNRAS.189...95P} Pagel, B.~E.~J., Edmunds,
M.~G., Blackwell, D.~E., Chun, M.~S., \& Smith, G.\ 1979, \mnras, 189, 95

\bibitem[Pagel, Simonson, Terlevich, \& Edmunds(1992)]{1992MNRAS.255..325P}
Pagel, B.~E.~J., Simonson, E.~A., Terlevich, R.~J., \& Edmunds, M.~G.\ 1992,
\mnras, 255, 325

\bibitem[Peimbert(2002)]{2002RMxAC..12..275P} Peimbert, M.\ 2002, Revista 
Mexicana de Astronomia y Astrofisica Conference Series, 12, 275 

\bibitem[Pilyugin(2000)]{2000A&A...362..325P} Pilyugin, L.~S.\ 2000, \aap, 362,
325

\bibitem[Pilyugin(2001)]{2001A&A...374..412P} Pilyugin, L.~S.\ 2001, \aap, 374,
412

\bibitem[Popescu, Hopp, \& Rosa(1999)]{1999A&A...350..414P} Popescu, C.~C.,
Hopp, U., \& Rosa, M.~R.\ 1999, \aap, 350, 414

\bibitem[Richer, McCall, \& Stasinska(1998)]{1998A&A...340...67R} Richer, M.,
McCall, M.~L., \& Stasinska, G.\ 1998, \aap, 340, 67

\bibitem[]{} Sandage, A.\ 1961, Hubble Atlas of Galaxies, Carnegie Institution
of Washington Publication 618

\bibitem[]{} Sandage, A., \& Tammann, G.~A.\ 1981, Revised Shapley-Ames Catalog
of Bright Galaxies, Carnegie Inst.~of Washington, Publ.~635

\bibitem[Sandage \& Binggeli(1984)]{1984AJ.....89..919S} Sandage, A.~\&
Binggeli, B.\ 1984, \aj, 89, 919. 

\bibitem[Shaw \& Dufour(1995)]{1995PASP..107..896S} Shaw, R.~A.~\& Dufour,
R.~J.\ 1995, \pasp, 107, 896

\bibitem[Sinclair (1996)]{1996 LPTN} Sinclair,~J.~1996, La Palma Tech. Note, 100

\bibitem[Skillman(1989)]{1989ApJ...347..883S} Skillman, E.~D.\ 1989, \apj, 
347, 883 

\bibitem[Skillman, Kennicutt, \& Hodge(1989)]{1989ApJ...347..875S} Skillman,
E.~D., Kennicutt, R.~C., \& Hodge, P.~W.\ 1989, \apj, 347, 875

\bibitem[Solanes et al.(2001)]{2001ApJ...548...97S} Solanes, J.~, Manrique,
A., Garc{\' \i }a-G{\' o}mez, C., Gonz{\' a}lez-Casado, G., Giovanelli, R., \&
Haynes, M.~P.\ 2001, \apj, 548, 97

\bibitem[Stasinska(1999)]{1999...} Stasinska, G.\ 1999, in ``Dwarf 
Galaxies and Cosmology'', ed. T.X. Thuan, C. Balkowsky, V. Cayatte, J.T. 
Than Van (Editions Frontieres), 259 

\bibitem[Stasi{\' n}ska, Schaerer, \& Leitherer(2001)]{2001A&A...370....1S} 
Stasi{\' n}ska, G., Schaerer, D., \& Leitherer, C.\ 2001, \aap, 370, 1 

\bibitem[Sutherland \& Dopita(1993)]{1993ApJS...88..253S} Sutherland, 
R.~S.~\& Dopita, M.~A.\ 1993, \apjs, 88, 253 

\bibitem[van Zee, Salzer, \& Haynes(1998)]{1998ApJ...497L...1V} van Zee, 
L., Salzer, J.~J., \& Haynes, M.~P.\ 1998, \apjl, 497, L1 

\bibitem[van Zee et al.(1998)]{1998AJ....116.2805V} van Zee, L., Salzer, J.~J.,
Haynes, M.~P., O'Donoghue, A.~A., \& Balonek, T.~J.\ 1998, \aj, 116, 2805

\bibitem[Vigroux, Lachieze-Rey, Thuan, \& Vader(1986)]{1986AJ.....91...70V}
Vigroux, L., Lachieze-Rey, M., Thuan, T.~X., \& Vader, J.~P.\ 1986, \aj, 91, 70

\bibitem[V{\' i}lchez(1995)]{1995AJ....110.1090V} V{\' \i}lchez, J.~M.\ 1995,
\aj, 110, 1090

\bibitem[V{\' i}lchez(1999)]{1999cezh.conf..175V} V{\' \i}lchez, J.~M.\ 1999, in
Chemical Evolution from Zero to High Redshift , 175

\bibitem[V{\' i}lchez \& Esteban(1996)]{1996MNRAS.280..720V} V{\' \i}lchez,
J.~M.~\& Esteban, C.\ 1996, \mnras, 280, 720

\bibitem[Whitford(1958)]{1958AJ.....63..201W} Whitford, A.~E.\ 1958, \aj, 63,
201

\bibitem[Zaritsky, Kennicutt, \& Huchra(1994)]{1994ApJ...420...87Z} 
Zaritsky, D., Kennicutt, R.~C., \& Huchra, J.~P.\ 1994, \apj, 420, 87 

\end{thebibliography}
\end{document}